\documentclass[IEEE journal, lettersize]{IEEEtran}
\usepackage{algorithmic}
\usepackage{url}

\usepackage{optidef} 
\usepackage{cite}
\usepackage{amsmath,amssymb,amsfonts}
\usepackage{bm}
\usepackage{algorithm}
\usepackage{algorithmic}
\usepackage{float}
\usepackage{amsthm}
\usepackage{graphics}
\usepackage{graphicx}
\usepackage{textcomp}
\usepackage{xcolor}
\usepackage{dsfont}
\usepackage{comment}
\usepackage{rotating}
\usepackage{makecell}
\usepackage{multirow}

\theoremstyle{plain}

\def\BibTeX{{\rm B\kern-.05em{\sc i\kern-.025em b}\kern-.08em
		T\kern-.1667em\lower.7ex\hbox{E}\kern-.125emX}}
\usepackage{subcaption}    
\definecolor{mygreen}{rgb}{0.01, 0.75, 0.24}    
\usepackage{arydshln}

\begin{document}

\title{Energy Saving in Cell-Free Massive MIMO ISAC for Ultra-Reliable Target-Aware Actuation }

\author{\IEEEauthorblockN{
Zinat Behdad\IEEEauthorrefmark{1}, \"Ozlem Tu\u{g}fe Demir\IEEEauthorrefmark{2}, Ki Won Sung\IEEEauthorrefmark{1}, and Cicek Cavdar\IEEEauthorrefmark{1}
\thanks{This work has been part of Celtic-Next project RAI-6Green: Robust and AI Native 6G for Green Networks with project-id: C2023/1-9 and 6G-SUSTAIN: Sensing Integrated Elastic 6G Networks for Sustainability. Both projects are funded by Vinnova in Sweden. \"O. T. Demir was supported by 2232-B International Fellowship for Early Stage Researchers Programme funded by the Scientific and Technological Research Council of T\"urkiye.}}

\IEEEauthorblockA{\IEEEauthorrefmark{1}Department of Communication Systems, KTH Royal Institute of Technology, Stockholm, Sweden 
(\{zinatb, sungkw, cavdar\}@kth.se)}\\
\IEEEauthorblockA{\IEEEauthorrefmark{2}Department of Electrical and Electronics Engineering, Bilkent University, Ankara, Turkiye  (ozlemtugfedemir@bilkent.edu.tr)}
}



\maketitle

\begin{abstract}
Emerging 6G sensing-based applications rely on ultra-reliable target-aware actuation, where timely and accurate sensing information triggers critical actions. Achieving this requires tightly integrated sensing and communication (ISAC) under stringent reliability and latency constraints. This paper investigates ISAC in a downlink cell-free massive multiple-input multiple-output (CF-mMIMO) system supporting multi-static sensing and ultra-reliable low-latency communications (URLLC). We propose a joint power and blocklength allocation algorithm to minimize total energy consumption, accounting for both radio-site and cloud-side processing energy for communication and sensing, while meeting communication and sensing requirements. The non-convex optimization problem is solved using a combination of feasible point pursuit–successive convex approximation (FPP-SCA), concave-convex programming (CCP), and fractional programming techniques. We consider two types of target detectors: clutter-aware and clutter-unaware, each with distinct complexity and performance trade-offs. A computational complexity analysis based on giga-operations per second (GOPS) is conducted to quantify the processing requirements of communication and sensing tasks. We also introduce the refreshing rate for sensing information and derive a closed-form expression that accounts for both observation and processing delays. Simulation results show that the proposed algorithm achieves up to 34\% energy reduction compared to schemes using the maximum allowable blocklength and enhances detection capability while consuming less total energy. Clutter-aware detectors, despite higher complexity, require fewer antennas and sensing receive APs, yielding up to $40\%$ energy savings.

\end{abstract}


\section{Introduction}
\IEEEPARstart{S}{ix}-generation (6G) mobile networks are expected to offer various sensing-based applications such as autonomous vehicles, smart homes/cities/factories, remote healthcare, industrial Internet-of-things (IIoT), and robot control in target/environment-aware scenarios \cite{liu2022integrated,zhou2022integrated}. These applications introduce ultra-reliable target-aware actuation use cases, in which a communication and sensing system must detect the presence, position, and motion of a target (e.g., a human or object) with high accuracy and minimal delay, and then trigger an actuation response (e.g., controlling a robot, vehicle, or machine) with guaranteed reliability and tightly bounded latency.

To support such use cases, the system must obtain and deliver sensing information to user equipment (UE) with a minimum reliability of 99.999\% and an end-to-end (E2E) latency of less than 10–150\,ms \cite{salehi2023URLLC, peng2023resource, ding2022joint}.
This demand necessitates two key aspects in 6G networks: integrated sensing and communication (ISAC) and ultra-reliable low-latency communication (URLLC). 

In ultra-reliable target-aware actuation use cases -- such as traffic control and autonomous vehicles -- real-time transmission of sensing information to URLLC UEs is essential, as the timeliness of updates directly affects the performance and safety. To characterize how frequently sensing information must be refreshed, this paper adopts the concept of the “refreshing rate”, defined by 3GPP as the number of sensing measurements and updates delivered per unit time \cite{3gpp_ts_22_137}. In ISAC systems, sensing and communication are executed jointly within each transmission, which makes the refreshing rate inherently dependent on the communication blocklength. Longer blocklengths reduce the refreshing rate since each joint sensing–communication operation takes longer to complete. This trade-off highlights the potential for optimizing blocklength to achieve a balance between reliable communication and a high refreshing rate.

Although URLLC and ISAC have traditionally been studied as separate domains, emerging target-aware actuation applications demand a unified design. A joint optimization framework is therefore necessary to simultaneously satisfy the stringent requirements of both communication reliability and sensing timeliness, thereby fully realizing the integration benefits of ISAC-enabled URLLC systems.

In this context, cell-free massive multiple-input multiple-output (CF-mMIMO) has emerged as a strong candidate for meeting the stringent requirements of URLLC applications \cite{lancho2023cell, ren2020joint, nasir2021cell}. It offers high reliability by compensating for large path-loss variations and significantly improves the performance of cell-edge UEs. Moreover, CF-mMIMO is well-suited for implementing ISAC, as its distributed access points (APs) and their coordination can support bi-static and multi-static sensing configurations \cite{sakhnini2022target, behdad2022power, demirhan2023cell, buzzi2024scalability}. These configurations eliminate the need for full-duplex capabilities at individual APs.

While CF-mMIMO networks offer significant performance advantages, integrating sensing functionality is anticipated to substantially increase transmission energy consumption, resulting in up to a tenfold reduction in energy efficiency \cite{behdad2023URLLC}. This challenge is further compounded by the fact that the combined demands of baseband processing and sensing operations may result in higher processing energy consumption than in conventional communication-only systems. Most existing research on green CF-mMIMO without sensing integration has primarily focused on optimizing transmit power, with relatively limited attention given to processing energy consumption, aside from a few notable exceptions \cite{demir2022cell, demir2023cell}. Neglecting processing energy consumption may become increasingly inaccurate in future deployments, since CF-mMIMO systems are envisioned to employ a large number of low-cost APs together with centralized cloud-based processing, which reduces the relative contribution of radio-frequency hardware while increasing the importance of computational energy consumption.
Moreover, the impact of sensing functionalities on processing energy consumption in such networks remains insufficiently explored. 

 It is important to note that communication system efficiency is typically measured in terms of data rate, while sensing tasks are evaluated using task-specific metrics, such as the refreshing rate, which quantifies the number of successful executions of a sensing task. Completing a sensing task involves not only transmitting signals and receiving reflections from the target but also performing extensive post-reception processing to extract the desired information. This post-reception processing is a fundamental component of the sensing operation and must be explicitly considered in system design and analysis. This distinction highlights the need for a comprehensive energy consumption analysis that accounts for both transmission and processing costs. 

Motivated by these considerations, this paper addresses the following key research question: How can sensing be effectively integrated into CF-mMIMO systems supporting URLLC, while minimizing total energy consumption?
\begin{figure}[t!]
\centerline{\includegraphics[trim={0mm 0mm 0mm 0mm},clip,
width=0.5\textwidth]{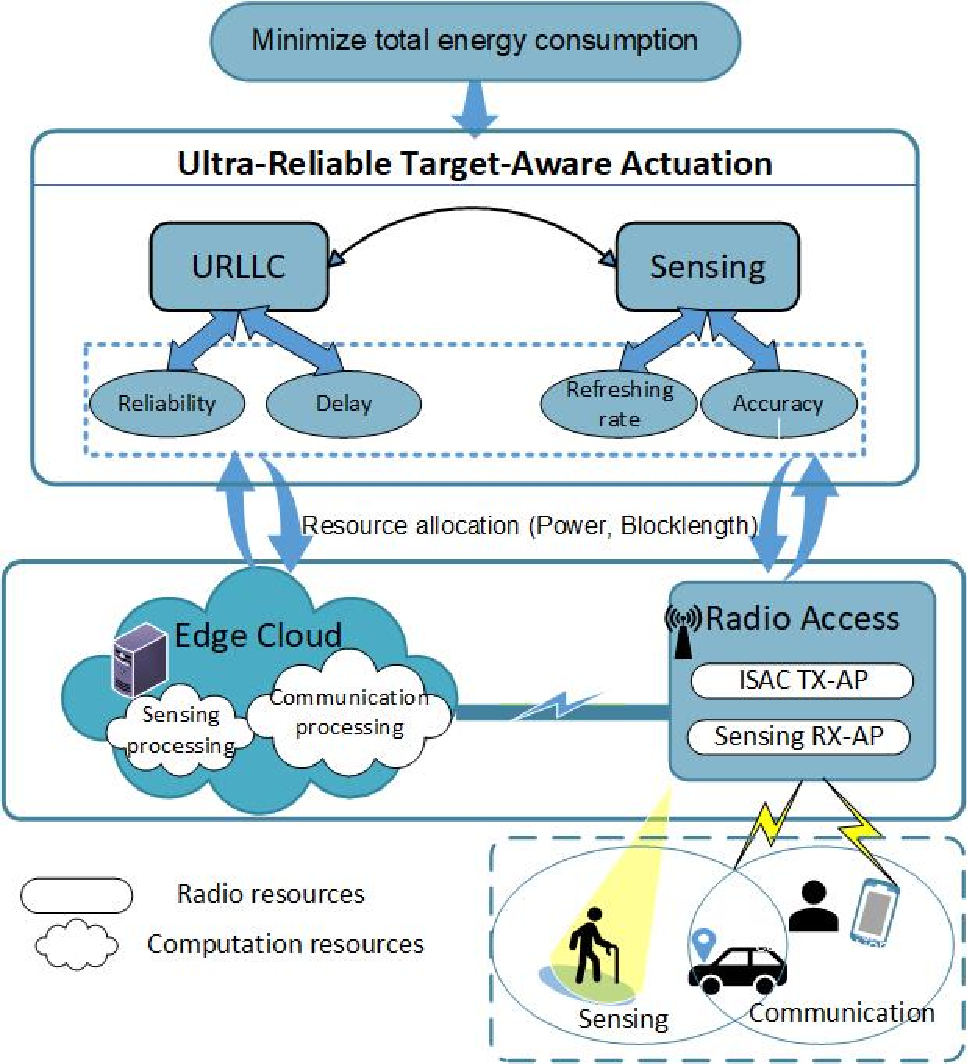}}
\caption{ISAC network architecture in CF-mMIMO With URLLC and joint resource allocation.}
\label{fig:architecture}\vspace{-4mm}
\end{figure}

Fig.~\ref{fig:architecture} illustrates the integration of communication and sensing functionalities in a CF-mMIMO network. The objective is to jointly optimize blocklength and power allocation to minimize total energy consumption across both the radio access and edge cloud domains. URLLC requirements--such as reliability and latency--and sensing requirements--such as refreshing rate and accuracy--are jointly addressed through coordinated resource allocation. The cloud manages centralized processing for both sensing and communication, while distributed APs enable multi-static sensing alongside communication services. This architecture leverages shared radio and computation resources to facilitate joint performance optimization across sensing and communication domains.
\vspace{-4mm}
\subsection{Literature Review}
The finite blocklength regime and URLLC have primarily been studied in the context of cellular networks (see \cite{nasir2020resource, soleymani2023optimization, van2022joint,ren2019joint, yang2024joint} and references therein), with limited research addressing URLLC in CF-mMIMO networks. 
For instance, although \cite{ren2019joint, yang2024joint} study joint power and blocklength optimization to enhance reliability in factory automation and heterogeneous networks, these results do not extend to CF-mMIMO.
The potentials of massive MIMO in meeting URLLC requirements for industrial automation are investigated in \cite{lancho2021cell, ostman2021urllc,alonzo2021cell}. 
Most of the works on CF-mMIMO and URLLC focus on data rate and energy efficiency maximization. 
The authors in \cite{elwekeil2022power}
propose two power optimization algorithms to provide URLLC for traditional ground UEs and uncrewed aerial vehicles (UAVs) in a CF-mMIMO system, with a focus on maximizing either the sum or the minimum URLLC rate. A power control algorithm based on geometric programming is proposed in \cite{PerformanceMIMO-URLLC} to enhance the downlink sum-rate. 
Max-min rate optimization and energy efficiency optimization are studied in \cite{nasir2021cell}, where the authors adapt a special class of conjugate beamforming for a CF-mMIMO with single-antenna APs. The authors in \cite{DeepEE} propose an iterative search-based two-stage energy-efficiency optimization algorithm and a convolutional neural network architecture for uplink power optimization, assuming a fixed blocklength.

The above-mentioned works do not consider sensing requirements. Indeed, there are a few works that jointly consider URLLC and ISAC.
In \cite{ding2022joint}, a joint precoding scheme is proposed to minimize transmit power, satisfying sensing and delay requirements. Moreover, joint ISAC beamforming and scheduling design is addressed in \cite{zhao2022jointbeam} and \cite{zhao2024joint} with a focus on the coexistence of periodic and aperiodic traffic to balance the trade-off between their corresponding performance. The aperiodic traffic is triggered by sensing information about the stochastic environment.

The consideration of E2E energy-awareness has been explored in various contexts, as reflected in prior works such as \cite{alabbasi2018optimal,masoudi2022energy,demir2022cell,demir2023cell, yang2019efficient}. In particular, \cite{demir2023cell} studied fully virtualized E2E power minimization problem for CF-mMIMO on O-RAN architecture by taking the radio, fronthaul, and processing resources into account. 
Joint UE scheduling and power allocation scheme for cell-free ISAC systems is studied in \cite{cao2023joint}, which aims to maximize the sum rate of the communication UEs and the sensing targets. However, the authors do not consider either URLLC requirements or energy minimization.

\begin{table*}[t]
    \centering
    \caption{Novelty Comparison of the Proposed Work with Existing Literature.}
\label{tab:novelty_comparison}
    
    \centering
    \begin{tabular}{|c|c|c|c|c|c|c|c|}
    \hline
      \multirow{1.5}{*}{ Work}&\multirow{1.5}{*}{ISAC}&  \multirow{1.5}{*}{URLLC}& \multirow{1.5}{*}{ MIMO}& \multirow{1.5}{*}{ Cell-free}&  \multirow{1.5}{*}{Target detection}& \multirow{1.5}{*}{\shortstack{Processing \&\\radio energy}}& \multirow{1.5}{*}{\shortstack{Blocklength\\ optimization}}\\[4mm]
      \hline
      \cite{ding2022joint} & \checkmark & \checkmark &  &  &  &  &  \\[2mm]
      \hdashline
      \cite{nasir2021cell} &  & \checkmark & \checkmark & \checkmark &  & \checkmark &  \\[2mm]
      \hdashline
      \cite{behdad2023URLLC} & \checkmark & \checkmark & \checkmark & \checkmark & \checkmark &  &  \\[2mm]
      \hdashline
      \cite{nasir2020resource, soleymani2023optimization, van2022joint} &  & \checkmark &  &  &  &  &  \\[2mm]
      \hdashline
		\cite{ren2019joint,yang2024joint} &  & \checkmark &  &  &  &  &\checkmark  \\[2mm]
      \hdashline
      \cite{lancho2021cell, ostman2021urllc,alonzo2021cell} &  & \checkmark & \checkmark &  &  &  &  \\[2mm]
      \hdashline
    \cite{elwekeil2022power,PerformanceMIMO-URLLC,DeepEE} &  & \checkmark & \checkmark & \checkmark &  &  &  \\[2mm]
      \hdashline      
      \cite{zhao2022jointbeam,zhao2024joint} & \checkmark & \checkmark & \checkmark &  &  &  &  \\[2mm]
      \hdashline
      \multirow{2}{*}{\shortstack{\cite{demir2022cell,demir2023cell},\\ \cite{alabbasi2018optimal,masoudi2022energy, yang2019efficient} }} 
        &  &  & \checkmark & \checkmark &  & \checkmark &  \\[5mm]
      \hdashline
      This paper & \checkmark & \checkmark & \checkmark & \checkmark & \checkmark & \checkmark & \checkmark \\[2mm]
      \hline
    \end{tabular}\vspace{-4mm}
\end{table*}
Table~\ref{tab:novelty_comparison} presents a comparison highlighting the novelty of this work relative to existing studies. To the best of our knowledge, our previous work \cite{behdad2023URLLC} was the first to investigate ISAC in CF-mMIMO systems with URLLC UEs from an energy-efficiency perspective.
However, that work assumed a fixed blocklength and did not account for cloud-side processing energy. In contrast, this paper incorporates finite-blocklength optimization and giga-operations per second (GOPS)-based processing-energy modeling, enabling a joint radio–cloud energy optimization framework that reveals new trade-offs between communication reliability, sensing refreshing rate, and processing workload.

Although finite-blocklength communication analysis, GOPS-based processing models, and sensing refreshing rate definitions have been studied separately in prior work, their interaction in CF-mMIMO ISAC systems remains largely unexplored. 
In this paper, we address this gap by formulating a joint blocklength and power optimization problem that minimizes total energy consumption, accounting for energy consumption at distributed radio APs and at the centralized cloud. The blocklength affects reliability, sensing timeliness, and cloud processing workload, revealing a cross-layer coupling between radio transmission and cloud computation. 

\vspace{-3mm}
\subsection{Contributions}\vspace{-2mm}
This paper develops a unified framework that explicitly captures the above-mentioned cross-layer coupling and exploits it for joint radio--cloud energy optimization under URLLC and sensing constraints. 
The main contributions of this paper are outlined as follows:
\begin{itemize}
    \item We incorporate the sensing refreshing rate, defined in 3GPP specifications, into the system-level ISAC framework and derive a closed-form expression that captures both sensing observation time and cloud processing delay. 
    \item We analyze the URLLC and sensing requirements in CF-mMIMO ISAC systems and show that the blocklength simultaneously determines communication reliability, sensing refreshing rate, and the cloud processing workload. By jointly considering transmission delay, processing delay, and refreshing rate constraints, we establish minimum and maximum tolerable blocklength that satisfy these requirements. This reveals a previously unexplored coupling between radio transmission parameters and cloud computational energy consumption.
    \item We develop a unified energy model that jointly captures radio transmission energy and cloud-side processing energy required for both communication and sensing tasks. By quantifying computational complexity in terms of GOPS, the model explicitly incorporates sensing processing complexity into the system-level energy optimization framework.
    \item We formulate a joint blocklength and transmit-power optimization problem that minimizes the total energy consumption of CF-mMIMO ISAC systems, accounting for both radio transmission energy and cloud processing energy. The problem incorporates URLLC reliability, transmission delay, sensing SINR, sensing refreshing rate, and processing delay constraints, which jointly determine the minimum and maximum feasible blocklength. The resulting non-convex problem is solved using a tractable approach based on feasible point pursuit–successive convex approximation (FPP-SCA), concave–convex programming (CCP), and fractional programming.
    \item We evaluate two target detection schemes based on the maximum a posteriori ratio test (MAPRT) and quantify the effect of their computational complexity on cloud processing energy and overall system energy consumption. Numerical results show that sensing processing can dominate the total energy consumption and that joint blocklength–power optimization can achieve up to 34\% energy savings. The results also highlight design trade-offs among detector complexity, antenna deployment, and sensing-receiver activation.
\end{itemize}

The rest of the paper is organized as follows: 
Section~\ref{sec:system_model} introduces the system model. Section~\ref{sec:URLLC_analysis} provides URLLC analysis, considering decoding error probability (DEP) and delay in the finite blocklength regime. Section~\ref{sec:sensing_analysis} covers the sensing analysis. Section~\ref{sec:power_model} describes the total power consumption model and derives GOPS analysis for both communication and sensing. Optimization problems are presented in Section~\ref{sec:optimization}, followed by numerical results and conclusions in Sections~\ref{sec:results} and \ref{sec:conclusion}, respectively.

\emph{Notations}: Scalars, vectors, and matrices are denoted by regular font, boldface lowercase, and boldface uppercase letters, respectively. The superscripts $(\cdot)^T$, $(\cdot)^*$, and $(\cdot)^H$ show the transpose operation, complex conjugate, and Hermitian transpose, respectively. The diagonalization and the block diagonalization operations are denoted by $\mathrm{diag}(\cdot)$ and $\mathrm{blkdiag}(\cdot)$, respectively. 
The trace and real parts of a matrix are represented by $\mathrm{tr}(\cdot)$ and $\Re(\cdot)$, respectively. $\textbf{A}\otimes\textbf{B}$ represents the Kronecker product between matrix $\textbf{A}$ and $\textbf{B}$. The absolute value of a scalar is denoted by $\vert \cdot \vert$ while $\Vert \cdot \Vert$ shows the Euclidean norm of a vector and $\mathbb{E}\left\{.\right\}$ denotes the expected value.

\vspace{-4mm}
\section{System Model}
\label{sec:system_model}
\begin{figure}[tbp]
\centerline{\includegraphics[trim={0mm 0mm 0mm 0mm},clip,
width=0.32\textwidth]{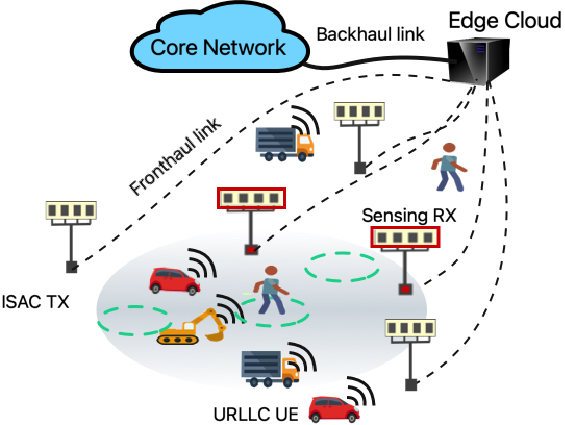}}
\caption{ISAC system model in CF-mMIMO with URLLC UEs.}
\label{fig1}\vspace{-4mm}
\end{figure}

We study ISAC in a CF-mMIMO system under URLLC scenarios and multi-static sensing, as shown in Fig.~\ref{fig1}. 
There are $N_{\rm tx}$ ISAC  transmit APs, $N_{\rm rx}$ sensing receive APs,  $N_{\rm ue}$ URLLC UEs and one candidate sensing location. Each AP is equipped with an array of $M$ antennas configured in a horizontal uniform linear array (ULA) with half-wavelength spacing. The respective array response vector is $
       \textbf{a}(\varphi,\vartheta) =\begin{bmatrix}
          1\!&\! e^{j\pi \sin(\varphi)\cos(\vartheta)}& \ldots& e^{j(M-1)\pi\sin(\varphi)\cos(\vartheta)}
        \end{bmatrix}^T,$
 where $\varphi$ and $\vartheta$ are the azimuth and elevation angles from the AP to the target location, respectively \cite{bjornson2017massive}.

We consider the original form of CF-mMIMO \cite{ lancho2023cell}, in which all ISAC transmit APs jointly serve the URLLC UEs by transmitting centralized precoded signals containing both communication and sensing symbols.\footnote{Note that scalable centralized implementations can be achieved through user-centric clustering, where each AP serves only a limited number of UEs\cite{cell-free-book}.} At the same time, the sensing receive APs simultaneously sense the candidate location.   
 
The APs operate in finite blocklength regime, where a packet of $b_i$ bits is sent to UE $i$ within a transmission block of length $L=L_p+L_d$ symbols over a coherence bandwidth $B$. Here, $L_p$ and $L_d$ denote the number of pilots and data symbols, respectively. Each transmission has a duration $T=\frac{L}{B}$, which is usually expected to be shorter than one coherence time $T_ c$, i.e., $T<T_ c$ \cite{nasir2021cell}. 
For simplicity and without loss of generality, we assume that each transmission block includes a channel estimation phase.
\subsection{Downlink ISAC Transmission}
Let $s_i[m]$ and $s_0[m]$ denote the downlink communication symbol for UE $i$ and the sensing symbol at time instance $m$, respectively. The symbols are independent, zero-mean, and unit-power. In addition, let $\rho_i\geq 0$ and $\rho_0\geq 0$ represent the power control coefficients assigned to UE $i$ and to the sensing signal, respectively. 
The transmitted signal from transmit AP $k\in\{1, \ldots, N_{\rm tx}\}$ at time instance $m$ is given by
\begin{equation}\label{x_k}
    \textbf{x}_k[m]= \sum_{i=0}^{N_{\rm ue}} \sqrt{\rho_{i}}\textbf{w}_{i,k} s_{i}[m]=\textbf{W}_k \textbf{D}_{\rm s}[m]\boldsymbol{ \rho},
\end{equation}
where $\textbf{w}_{i,k}\in \mathbb{C}^{M}$ and $\textbf{w}_{0,k}\in \mathbb{C}^M$ denote the transmit precoding vectors at AP $k$ for UE $i$ and the sensing signal, respectively.
For notational convenience, we define $\textbf{W}_k= \begin{bmatrix}
\textbf{w}_{0,k} & \textbf{w}_{1,k} & \cdots & \textbf{w}_{N_{\rm ue},k}
\end{bmatrix}$ as the matrix that collects the precoding vectors at transmit AP $k$. The diagonal matrix  ${\textbf{D}}_{\rm s}[m]=\mathrm{diag}\left(s_0[m],s_1[m],\ldots,s_{N_{\rm ue}}[m]\right)$ contains the sensing and communication symbols, and the vector $\boldsymbol{\rho}=[\sqrt{\rho_0} \ \ldots \ \sqrt{\rho_{N_{\rm ue}}}]^T$ is the corresponding power-control coefficients.

The communication channels are modeled as spatially correlated Rician fading, assumed to remain constant during each coherence block, and the channel realizations are independent of each other. Let $\textbf{h}_{i,k}\in \mathbb{C}^M$ denote the channel between ISAC AP $k$ and UE $i$, modeled as
\begin{align}
    \textbf{h}_{i,k} =e^{j\varphi_{i,k}}\bar{\textbf{h}}_{i,k}\,+\,\tilde{\textbf{h}}_{i,k}, \label{eq:channel}
\end{align}
which consists of a semi-deterministic line-of-sight (LOS) path, represented by $e^{j\varphi_{i,k}}\bar{\textbf{h}}_{i,k}$ with unknown phase-shift $\varphi_{i,k}\sim \mathcal{U}[0,2\pi)$, i.e., uniformly distributed on $[0,2\pi)$, and a stochastic non-LOS (NLOS) component $\tilde{\textbf{h}}_{i,k}\sim \mathcal{CN}(\textbf{0},\textbf{R}_{i,k})$ with the spatial correlation matrix $\textbf{R}_{i,k}\in \mathbb{C}^{M\times M}$. Both $\bar{\textbf{h}}_{i,k}$ and $\textbf{R}_{i,k}$ include the combined effect of geometric path loss and shadowing. In the considered system, the communication channels are estimated at the CPU using linear minimum mean-squared error (LMMSE) channel estimation based on uplink pilot signaling, and the resulting channel estimates are used for precoding design. 
We concatenate the channel vectors $\textbf{h}_{i,k}$ for UE $i$ in the collective channel vector $\textbf{h}_{i}$ as
\begin{align}
   \textbf{h}_{i}=\begin{bmatrix}
\textbf{h}_{i,1}^T& \ldots&
\textbf{h}_{i,N_{\rm tx}}^T
\end{bmatrix}^T\in \mathbb{C}^{N_{\rm tx}M} ,
\end{align}
 
The received signal at UE $i$ is given as

\begin{align}    \label{y_i-2}
     y_i[m] =&\underbrace{\sqrt{\rho_i}\textbf{h}_{i}^{H}\textbf{w}_{i} s_{i}[m]}_{\textrm{Desired signal}}+ \underbrace{\sum_{j=1,j\neq i}^{N_{\rm ue}}\sqrt{\rho_j}\textbf{h}_{i}^{H}\textbf{w}_{j} s_{j}[m]}_{\textrm{Interference signal due to other UEs}}\nonumber\\
     &+ \underbrace{\sqrt{\rho_0}\textbf{h}_{i}^{H}\textbf{w}_{0} s_{0}[m]}_{\textrm{Interference signal due to sensing}}+ \underbrace{n_i[m]}_{\textrm{Noise}},
\end{align}
where $n_i[m]\!\sim\!\mathcal{CN}(0,\sigma_n^2)$ is the independent receiver noise at UE $i$ at time instance $m$. The concatenated  vectors 
$\textbf{w}_i=\left[
        \textbf{w}_{i,1}^T \ \textbf{w}_{i,2}^T \ \hdots \ \textbf{w}_{i,N_{\rm tx}}^T
    \right]^T\in \mathbb{C}^{N_{\rm tx}M},
$ and
$
\textbf{w}_0=\left[\textbf{w}_{0,1}^T \ \textbf{w}_{0,2}^T \ \hdots \ \textbf{w}_{0,N_{\rm tx}}^T\right]^T\in\mathbb{C}^{N_{\rm tx}M}$
are the centralized unit-norm precoding vectors obtained based on regularized zero forcing (RZF) and zero forcing (ZF) approaches, respectively. 

The unit-norm RZF precoding vector for UE $i$ is given as $\textbf{w}_{i}=\frac{\bar{\textbf{w}}_{i}}{\left \Vert \bar{\textbf{w}}_{i}\right \Vert}$, with

\begin{equation} \label{wi}
     \bar{\textbf{w}}_{i}
     \!=\!\left(\sum\limits_{j=1}^{N_{\rm ue}}\hat{\textbf{h}}_j\hat{\textbf{h}}_j^H+\delta\textbf{I}_{N_{\rm tx}M}\right)^{\!\!-1}\!\!\hat{\textbf{h}}_i, \quad i=1,\ldots,N_{\rm ue},  
\end{equation}
where $\delta$ is the regularization parameter, and $\hat{\textbf{h}}_{j}=\begin{bmatrix}
\hat{\textbf{h}}_{j,1}^T& \ldots&
\hat{\textbf{h}}_{j,N_{\rm tx}}^T
\end{bmatrix}^T\in \mathbb{C}^{N_{\rm tx}M}$ 
is the LMMSE channel estimate of the communication channel $\textbf{h}_{j}$, obtained as in \cite{wang2020uplink}\footnote{ We omit the
explanations from this paper due to the space limitation.}. 
If the number of UEs is larger than the number of mutually orthogonal pilot sequences, then each pilot sequence may be assigned to multiple UEs using the pilot assignment algorithm in \cite[Algorithm 4.1]{cell-free-book}.

We aim to null the destructive interference from the sensing signal to the UEs by using the unit-norm ZF sensing precoding vector $\textbf{w}_{0} = \frac{\bar{\textbf{w}}_{0}}{\left \Vert \bar{\textbf{w}}_{0}\right \Vert}$, where 
   \begin{equation}\label{eq:w0}
       \bar{\textbf{w}}_{0}=\left(\textbf{I}_{N_{\rm tx}M}-\textbf{U}\textbf{U}^{H}\right)\textbf{h}_{0},
   \end{equation}
and $\textbf{U}$ is the unitary matrix with the orthogonal columns that span the column space of the matrix $\begin{bmatrix}
         \hat{\textbf{h}}_{1}&  \ldots& \hat{\textbf{h}}_{N_{\rm ue}}
        \end{bmatrix}$. $\textbf{h}_{0}= \begin{bmatrix}
\sqrt{\beta_1}\textbf{a}^T(\varphi_{1},\vartheta_{1})&  \ldots&
\sqrt{\beta_{N_{\rm tx}}}\textbf{a}^T(\varphi_{N_{\rm tx}},\vartheta_{N_{\rm tx}})
\end{bmatrix}^T\in \mathbb{C}^{N_{\rm tx}M}$ is the concatenated sensing channel between all the ISAC APs and the target, including the corresponding
channel gains $\beta_k$ and the array response vectors $\textbf{a}(\varphi_{k},\vartheta_{k})$ for $k=1,\ldots, N_{\rm tx}$.\footnote{CF-mMIMO systems with centralized processing and precoding typically provide abundant spatial degrees of freedom, since the total number of transmit antennas generally satisfies $M N_{\mathrm{tx}} \gg (N_{\mathrm{ue}} + 1)$.}

\vspace{-3mm}
\subsection{Multi-Static Sensing Reception}
We employ multi-static sensing in which multiple transmit and receive APs participate in target detection. 
A LOS connection is assumed to exist between the target location and each transmit and receive AP.
In the presence of the target, each receive AP observes a superposition of the desired target-reflected signals and undesired reflections from surrounding objects, commonly referred to as clutter. Since clutter is independent of the target’s presence, it is treated as interference from a sensing perspective. Without loss of generality, we assume that the LOS propagation paths between transmit and receive APs are known and can be effectively canceled, which is commonly adopted in cooperative ISAC and CF-mMIMO systems with centralized processing.\footnote{In practice, imperfect synchronization and calibration may lead to residual LOS interference, which can be modeled as additional interference that reduces the sensing signal-to-interference-plus-noise ratio (SINR). Recent works on distributed massive MIMO synchronization and calibration provide practical mechanisms, including advanced over-the-air techniques to mitigate such effects \cite{larsson2024massive, han2025network}.} As a result, the remaining interference consists solely of reflected paths caused by obstacles, which we refer to as target-free channels that are not known beforehand and can be inferred by the detector.

The target-free channel between transmit AP $k$ and receive AP $r$, denoted by $\textbf{H}_{r,k}\in \mathbb{C}^{M \times M}$, is a correlated Rayleigh fading channel that can be modeled as $\textbf{H}_{r,k} = \textbf{R}^{\frac{1}{2}}_{{\rm rx},(r,k)} \textbf{W}_{{\rm ch},{(r,k)}}\left(\textbf{R}^{\frac{1}{2}}_{{\rm tx},(r,k)}\right)^T$ using the Kronecker model \cite{Shiu2000a}. The matrix $\textbf{W}_{{\rm ch},(r,k)}\in \mathbb{C}^{M \times M}$ is a random matrix with independent and identically distributed (i.i.d.) entries following $\mathcal{CN}(0,1)$ distribution. The matrix $\textbf{R}_{{\rm rx},(r,k)} \in \mathbb{C}^{M \times M}$ represents the spatial correlation matrix at receive AP $r$ with respect to the direction of transmit AP $k$. Similarly, $\textbf{R}_{{\rm tx},(r,k)}\in \mathbb{C}^{M \times M}$ is the spatial correlation matrix at transmit AP $k$ with respect to the direction of receive AP $r$. The channel gain is determined by the geometric path loss and shadowing, and is included in the spatial correlation matrices.
    
The received signal at AP $r$ in the presence of the target is  
\begin{align}\label{y_rPrim}
          &\textbf{y}_r[m]
           =\underbrace{\sum_{k=1}^{N_{\rm tx}}\alpha_{r,k}\sqrt{\beta_{r,k} }\textbf{a}(\phi_{r},\theta_{r})\textbf{a}^{T}(\varphi_{k},\vartheta_{k})\textbf{x}_k[m]}_{\textrm{Desired reflections from  target}}\nonumber\\
           &+\underbrace{\sum_{k=1}^{N_{\rm tx}}\!\textbf{H}_{r,k}\textbf{x}_k[m]}_{\textrm{Clutter}}+\textbf{n}_r[m],\quad m\in\{1, \ldots, L_d\}
\end{align}
where $\textbf{n}_r[m]\sim \mathcal{CN}(\textbf{0},\sigma_n^2\textbf{I}_M)$ is the receiver noise at receive AP $r$.  
The angles $\phi_r$ and $\theta_r$ denote the azimuth and elevation angles from the target location to receive AP $r$, while $\varphi_k$ and $\vartheta_k$ represent the azimuth and elevation angles from transmit AP $k$ to the target location. 
The term $\alpha_{r,k}\!\sim\!\mathcal{CN}(0,\,1)$ represents the normalized RCS of the target for the corresponding path. We assume the RCS values are i.i.d. and follow the Swerling-I model, meaning that they are constant throughout the consecutive $L_d$ symbols collected for sensing \cite{richards2010principles}. The channel gain $\beta_{r,k}$ accounts for the path loss from transmit AP $k$ to receive AP $r$ via the target, as well as the bi-static radar cross-section (RCS) variance of the target, denoted by $\sigma_{\rm rcs}$. Following the radar range equation for bi-static sensing \cite[Chap.~2]{richards2010principles}, $\beta_{r,k}$ is computed as 
\begin{align} \label{eq:radar_range_equation}
    \beta_{r,k} = \frac{\lambda_c^2 \,\sigma_{\rm rcs}}{(4\pi)^3 d_{{\rm tx},k}^2 d_{{\rm rx},r}^2},
\end{align}
where $\lambda_c$ is the wavelength and $d_{{\rm tx},k}$ and $d_{{\rm rx},r}$ are distances from the target to transmit AP $k$ and receive AP $r$, respectively. 

For notational convenience, let ${\textbf{g}_{r,k}[m]\in \mathbb{C}^{M}}$ denote the known part of each reflected path in \eqref{y_rPrim}, defined as $\textbf{g}_{r,k}[m]\triangleq \sqrt{\beta_{r,k}}\textbf{a}(\phi_{r},\theta_{r})\textbf{a}^{T}(\varphi_{k},\vartheta_{k})\textbf{x}_k[m]$. We then form  $\textbf{G}_r[m] = \begin{bmatrix}\textbf{g}_{r,1}[m]& \ldots &\textbf{g}_{r,N_{\rm tx}}[m] \end{bmatrix}\in \mathbb{C}^{M \times N_{\rm tx}}$ and $\textbf{G}[m]=\mathrm{blkdiag}\left(\textbf{G}_{1}[m],\ldots,\textbf{G}_{N_{\rm rx}}[m]\right)$. We also define the concatenated transmit signal vector $\textbf{x}[m] = \begin{bmatrix}\textbf{x}^T_1[m]& \ldots &\textbf{x}^{T}_{N_{\rm tx}}[m] \end{bmatrix}^T\in \mathbb{C}^{ N_{\rm tx}M}$ and the matrix $\textbf{X}[m]=\left(\textbf{I}_{N_{\rm rx}} \otimes \left(\textbf{x}^T[m] \otimes \textbf{I}_{M}\right)\right)$, and let $\boldsymbol{\mathfrak{h}} \sim \mathcal{CN}\left(\textbf{0}, \textbf{R}\right)$ be the vectorized target-free channel with correlation matrix $\textbf{R}$\cite{behdad2023}. Moreover, let $\boldsymbol{\alpha}\in \mathbb{C}^{N_{\rm tx}N_{\rm rx}}$ be the concatenated vector of unknown RCS values. 
Each receive AP sends their respective signals to the cloud to form the concatenated received signal
    $\label{y-prime}
  \textbf{y}[m]= \begin{bmatrix}
    \textbf{y}_{1}^T[m]&  \ldots&\textbf{y}_{N_{\rm rx}}^T[m]
    \end{bmatrix}^T 
  $, given by 
\begin{align}\label{y_m}
     \textbf{y}[m]
     &=\textbf{G}[m] \boldsymbol{\alpha}+\textbf{X}[m] \boldsymbol{\mathfrak{h}}+\textbf{n}[m],
\end{align}
which is then processed in the cloud for target detection. 
\vspace{-3mm}
\subsection{Target Detection}\label{sec:detection}
Target detection is formulated as a binary hypothesis test as in~\eqref{hypothesis}, where a test statistic is computed at the cloud from the received signals $\textbf{y}[m]$ and compared against a threshold $\lambda_d$ that is determined by a desired false alarm probability threshold. 
\begin{align}\label{hypothesis}
   &\mathcal{H}_0 : \textbf{y}[m]= \textbf{X}[m] \boldsymbol{\mathfrak{h}}+\textbf{n}[m], \quad\hspace{13mm} m\in\{1,\cdots, L_d\} \nonumber\\
  &\mathcal{H}_1 :\textbf{y}[m]=\textbf{G}[m] \boldsymbol{\alpha}+\textbf{X}[m] \boldsymbol{\mathfrak{h}}+\textbf{n}[m], \quad m\in\{1,\cdots, L_d\}.
\end{align}
We employ two maximum a posteriori ratio test (MAPRT) detectors with two levels of complexity: the clutter-unaware detector and the clutter-aware detector, proposed in \cite{behdad2022power} and \cite[Lem. 2]{behdad2023}, respectively. The clutter-unaware detector ignores clutter to reduce complexity, whereas the clutter-aware detector accounts for unknown clutter. The corresponding test statistics are given by
\begin{align}
    &T_{\text{c-unaware}} =\textbf{a}^H \textbf{C}^{-1}\textbf{a},\label{eq:T_SP}
    \\
     &T_{\text{c-aware}} = \begin{bmatrix} \textbf{a}\\
     \textbf{b}
\end{bmatrix}^H\left( \begin{bmatrix}
 \textbf{C}& \textbf{E} \\ 
 \textbf{E}^H&\textbf{D} 
\end{bmatrix}^{-1}-\begin{bmatrix}
 \textbf{0}& \textbf{0} \\ 
 \textbf{0}&\textbf{D}^{-1} 
\end{bmatrix}\right)\begin{bmatrix}
\textbf{a}\\ 
\textbf{b}
\end{bmatrix}, \label{eq:T}
\end{align}
respectively, where

\begin{align}
    &\normalfont\textbf{a}= \sum_{m=1}^{L_d} \textbf{G}^H[m] \textbf{y}[m],\quad \normalfont\textbf{b}= \sum_{m=1}^{L_d} \textbf{X}^H[m] \textbf{y}[m],\label{eq:a-b}\\
    &\normalfont\textbf{C}= \sum_{m=1}^{L_d} \textbf{G}^H[m]\textbf{G}[m]+ \sigma_n^2 \textbf{I}_{N_{\rm rx}N_{\rm tx}}, \\ &\normalfont\textbf{D}= \sum_{m=1}^{L_d} \textbf{X}^H[m]\textbf{X}[m]+ \sigma_n^2\textbf{R}^{-1}, \label{eq:C-D}\\
   & \normalfont\textbf{E}= \sum_{m=1}^{L_d} \textbf{G}^H[m]\textbf{X}[m]. \label{eq:e-F}
\end{align}
Finally, the target is declared detected if the test statistic exceeds the threshold.    

\vspace{-2mm}
\section{Reliability and Delay Analysis for URLLC}\label{sec:URLLC_analysis}
\subsection{Reliability Analysis}
 First, we temporarily ignore the finite-blocklength regime. In the literature, an achievable spectral efficiency (SE) for the considered downlink system is typically derived under the assumption that UE $i$ only has access to the average effective channel gain $\mathbb{E}\left\{\textbf{h}_{i}^H \textbf{w}_{i}\right\}$. In this case, leveraging the channel hardening property of massive MIMO systems, the small-scale channel variations vanish, and the channel effectively behaves as deterministic from the UE perspective. Under this widely adopted assumption, and according to \cite[Thm.~6.1]{cell-free-book} and \cite[Lem.~1]{behdad2023}, the achievable SE can be expressed as
\begin{align}\label{E_sinr}
   \log_2{\left(1+ \overline{\mathsf{SINR}}^{(\rm dl)}_i\right)}
\end{align}
where $\overline{\mathsf{SINR}}^{(\rm dl)}_i$ denotes the effective SINR at UE $i$, given by 
\begin{align}\label{sinr_i}
&\overline{\mathsf{SINR}}^{(\rm dl)}_i=
    \frac{\rho_i\mathsf{b}_i^2}{\sum_{j=0}^{{\rm N_{ue}}}\rho_j\mathsf{a}_{i,j}^2+\sigma_n^2}, \quad i= 1,\ldots,N_{\rm ue}
    \end{align}
    with $ \mathsf{b}_i = \left \vert \mathbb{E}\left\{\textbf{h}_{i}^H \textbf{w}_{i}\right\}\right\vert$, $\mathsf{a}_{i,i} =\sqrt{\mathbb{E}\left\{\left \vert\textbf{h}_{i}^H \textbf{w}_{i}\right\vert^2\right\}- \mathsf{b}_i^2}$, and $
    \mathsf{a}_{i,j} = \sqrt{\mathbb{E}\left\{\left \vert\textbf{h}_{i}^H \textbf{w}_{j}\right\vert^2\right\}}$ for $j= 0,1,\ldots,N_{\rm ue}, \quad j\neq i$. The expectations are taken with respect to the random channel realizations. 

For most URLLC applications, short codewords are required to meet stringent latency constraints, typically corresponding to blocklengths of 50–400 symbols. However, operating in this regime increases the decoding error probability. Following \cite{ren2020joint}, the achievable rate of UE $i$ under finite blocklength can be approximated--under the assumption that average channel gains are known at the UEs--as
\begin{equation}\label{eq:R_i}
     R_i \! \approx \!
   \left(1\!-\!\beta\right)\! \log_2{\left(1+ \overline{\mathsf{SINR}}^{(\rm dl)}_i\right)}\!-\!\frac{Q^{-1}(\epsilon_{i})}{\ln(2)}\!\sqrt{\!\frac{(1\!-\!\beta)V_i}{L}},
\end{equation}
where $\beta = \frac{L_p}{L}$, $\epsilon_i$ denotes the DEP when transmitting $b_i$ bits to UE $i$,  $\overline{\mathsf{SINR}}^{(\rm dl)}_i$ is the effective downlink communication SINR for UE $i$,  $V_i= 1-\left(1+ \overline{\mathsf{SINR}}^{\rm (dl)}_{i}\right)^{-2}$ is the channel dispersion, and $Q(\cdot)$ refers to the Gaussian Q-function. Due to the fact that $V_i<1$, the following relationship holds 
\begin{align}\label{eq:R_i_lb}
     R_i & \!\geq \!   \left(1\!-\!\beta\right) \log_2{\left(1+ \overline{\mathsf{SINR}}^{(\rm dl)}_i\right)}-\!\frac{Q^{-1}(\epsilon_{i})}{\ln(2)}\sqrt{\frac{(1-\beta)}{L}}%
\end{align}
 Using \eqref{E_sinr}
and substituting $R_i = \frac{b_i}{L}$ into \eqref{eq:R_i_lb}, we obtain an upper bound for the DEP as
\begin{align}\label{upperbound}
    \epsilon_{i}^{\rm (ub)}\! \triangleq\!Q\!\left(\sqrt{L-L_p} \left[\ln\left(1+ \overline{\mathsf{SINR}}^{(\rm dl)}_i\right)-\frac{b_i \ln{2}}{L-L_p} \right]\right),
\end{align}
where $\epsilon_{i}\leq\epsilon_{i}^{\rm (ub)}$.
\vspace{-2mm}
\subsection{Delay Analysis}
The E2E delay experienced by UE $i$ can be decomposed into processing, transmission, propagation, and other network-related components, expressed as
\begin{align}
    D_i = D_i^{\rm p,c} + D_i^{\rm t} + D_i^{\rm p}+ D_i^{\rm p,u} + D_i^{\rm o},
\end{align}
where $D_i^{\rm p,c}$ denotes the processing delay at the edge cloud, $D_i^{\rm t}$ represents the wireless transmission delay, $D_i^{\rm p}$ is the propagation delay, $D_i^{\rm p,u}$ is the processing delay at the UE, and $D_i^{\rm o}$ captures additional delays such as queueing, and other network-related delays.

The transmission delay for each UE depends on the reliability requirement and the physical-layer transmission parameters and is given as
\begin{align}\label{eq:delay}
    D_i^{\rm t} = \frac{T}{1-\epsilon_i}
    = \frac{L}{B(1-\epsilon_i)}, \quad i = 1,\ldots,N_{\rm ue},
\end{align}
where $\frac{1}{1-\epsilon_i}$ denotes the average number of retransmissions. 

We assume that no service slicing is implemented in the system; thus, all UEs experience the same cloud-side processing delay, i.e., $D_i^{\rm p,c}=D^{\rm p,c}$. This delay depends on the computational complexity of the communication and sensing tasks as well as the computational resources available at the cloud. Let $G_{\rm cloud}$ be the total number of giga operations required for completing the communication and sensing tasks, as described in Section \ref{sec:power_model}, and $N_{\rm GPP}C_{\rm max}$ be the maximum computational capacity at the cloud, where $C_{\rm max}$ is the maximum processing capacity of each general-purpose processor (GPP) in GOPS and $N_{\rm GPP}$ is the number of GPPs. The processing delay at the cloud is given as \footnote{This linear formulation corresponds to an operating regime below capacity saturation, where queueing effects are negligible compared to deterministic service time \cite{harchol2013performance}.
In practice, such a regime can be ensured by allocating sufficient GPPs and optimizing the blocklength to keep the processing load below the saturation threshold.} 
\begin{align}\label{eq:D_pc}
    D^{\rm p,c}= \frac{G_{\rm cloud}}{N_{\rm GPP}C_{\rm max}}.
\end{align}

The remaining processing after reception occurs at the UE and depends on device-specific computational capabilities. The propagation delay and other network-related delays also depend on the UE and AP locations and fronthaul and backhaul specifications. Therefore, these delays are not explicitly modeled in this paper. Indeed, we focus on transmission and processing delays and leave the E2E delay analysis as future work.  In practice, the URLLC E2E delay requirement can be enforced by allocating a portion of the delay budget to propagation and other network-related delays, while the proposed framework meets the transmission and processing delay requirements. In delay-sensitive applications, the processing delay should usually be no greater than the duration of each transmission block, i.e., $ D^{\rm p,c}\leq T$ \cite{mao2016dynamic}, and the transmission delay should not exceed a threshold, denoted by $D_i^{(\rm th)}$.  
\vspace{-4mm}
\section{Sensing Analysis}\label{sec:sensing_analysis}
We assess the sensing performance in terms of detection probability, denoted by $P_{\rm d}$, under a certain false alarm probability $P_{\rm fa}$ as well as the refreshing rate $R_s$ defined as the number of sensing updates per time unit. Detection probability refers to the likelihood of correctly identifying the presence of a target, whereas false alarm probability denotes the likelihood of incorrectly detecting a target in the target's absence.

For a fixed false alarm probability, the detection probability is directly linked to the sensing SINR; thus, increasing the sensing SINR improves detection performance \cite[Chapters 3 and 15]{richards2010principles}. 
Consequently, a minimum sensing SINR threshold, denoted by $\gamma_{\rm s}$, should be ensured, i.e., $ \mathsf{SINR}_{\rm s}\geq \gamma_{\rm s}$. This requirement motivates the optimization problem formulated in Section~ \ref{sec:optimization}.\footnote{Higher-layer metrics, such as tracking and localization accuracy, inherently depend on detection reliability and refreshing rate; hence, improving these quantities directly enhances application-level performance. Stricter tail-latency constraints would mainly tighten the feasible region without altering the fundamental trade-offs.}  
The average sensing SINR is computed by taking the expectation with respect to the channel realizations and random symbols, given as 
\begin{align} \label{gamma_s_main}
     \overline{\mathsf{SINR}}_{\rm s}
     &= \frac{M \boldsymbol{\rho}^T \textbf{A}_D \boldsymbol{\rho}}{ MN_{\rm rx}\sigma_n^2+ \boldsymbol{\rho}^T \textbf{B}_D\boldsymbol{\rho} },
\end{align} 
where $\textbf{A}_D$ and $\textbf{B}_D$ are diagonal matrices with 
\begin{align}
    \left[\textbf{A}_D\right]_{ii} &= \sum_{r=1}^{N_{\rm rx}} \sum_{k=1}^{N_{\rm tx}} \beta_{r,k} \left( \textbf{W}_k^H \textbf{a}^{*}(\varphi_{k}, \vartheta_{k}) \textbf{a}^{T}(\varphi_{k}, \vartheta_{k}) \textbf{W}_k \right)_{ii}, \\
    \left[\textbf{B}_D\right]_{ii} &= \sum_{r=1}^{N_{\rm rx}} \sum_{k=1}^{N_{\rm tx}} \mathrm{tr} \left( \textbf{R}_{\mathrm{rx},(r,k)} \right) \left( \textbf{W}_k^H \textbf{R}^{T}_{{\rm tx},(r,k)} \textbf{W}_k \right)_{ii}.
\end{align}

Beyond the sensing SINR requirement, ISAC systems must also meet timing constraints associated with obtaining sensing information. In this context, the 3GPP specification \cite{3gpp_ts_22_137} defines the refreshing rate as the rate at which the sensing result is updated in the system.  
The refreshing rate $R_{\rm s}$ can be formulated as the inverse of the total time required to obtain the sensing information, given by
\begin{align}\label{eq:R_s}
    &R_{\rm s} =\frac{1}{T_{\rm obs}+D^{\rm p,c}
    } \quad(\rm updates/second),
\end{align} 
where $T_{\rm obs} =  T+ D^{\rm p}_{\rm 2-way}$ is observation time, representing the total duration of signal transmission and reception, and $D^{\rm p,c}$ is the processing time at the cloud, obtained from \eqref{eq:D_pc}. The observation time depends on the duration of each transmission and the maximum two-way propagation delay $D^{\rm p}_{\rm 2-way}$ corresponding to the path between the transmit AP, the target, and the receive AP. 

\vspace{-4mm}
\section{Joint Radio and Cloud Energy Consumption Model}\label{sec:power_model}
In this section, we first present a joint radio and cloud power consumption model that accounts for both the power consumed at the radio site, denoted by $P_{\rm radio}$, and the power consumed at the cloud, where all sensing and communication processing is performed, denoted by $P_{\rm cloud}$. Then, we present a closed-form expression of the total energy model.  
Total power consumption, taking into account both communication and sensing, is given as
\begin{align}\label{eq:Ptotal1}
    &P_{\rm total} = \underbrace{\Delta ^{\rm tr}P_{\rm tr}+\sum_{k=1}^{N_{\rm tx}} P_{\textrm{AP},0}^{\rm tx}+ \sum_{r=1}^{N_{\rm rx}} P_{\textrm{AP},0}^{\rm rx}
    }_{\triangleq P_{\rm radio}}\nonumber\\
    &\hspace{3mm}+ \underbrace{P_{\rm fixed} +\frac{1}{\sigma_{\rm cool}}\Bigg(N_{\rm GPP} P_{\rm cloud,0}^{\rm proc}+\Delta_{\rm cloud}^{\rm proc}
    \frac{C_{\rm cloud}}{C_{\rm max}} \Bigg)}_{\triangleq P_{\rm cloud}},
\end{align}
where $P_{\rm tr}=\sum_{j=0}^{N_{\rm ue}}\rho_j= \Vert \boldsymbol{\rho} \Vert^2$ is the total transmit power with $\boldsymbol{\rho}=[\sqrt{\rho_0} \ \ldots \ \sqrt{\rho_{N_{\rm ue}}}]^T$, $\Delta ^{\rm tr}$ is the slope of load-dependent power consumption at the APs. The terms $P_{\textrm{AP},0}^{\rm tx}$ and $P_{\textrm{AP},0}^{\rm rx}$ are the static power consumption of the ISAC transmit AP and the sensing receive AP, respectively.
 $P_{\rm fixed}$ is the fixed power consumption  
at the cloud, which is independent of the load, 
and $P_{\rm cloud,0}^{\rm proc}$ is the processing power consumption in the idle mode. $\sigma_{\rm cool}\in (0,1]$ and $\Delta_{\rm cloud}^{\rm proc}$ denote the cooling efficiency of the cloud and the slope of the load-dependent power consumption for cloud processing, respectively. 
Moreover, $C_{\rm cloud}$ is the total processing resource utilization in GOPS\cite{demir2023cell,masoudi2020cost}.\footnote{The adopted linear GOPS-based processing power model represents a first-order abstraction; although practical cloud platforms may exhibit nonlinear power behavior due to CPU utilization, queueing, and virtualization overheads, the key qualitative insights of the proposed framework remain unchanged.} The processing resource utilization is given as 
\begin{equation}
    C_{\rm cloud}= \frac{B}{L}G_{\rm cloud}=C_{\rm proc}^{\rm c}+C_{\rm proc}^{\rm s}
\end{equation} 
where $C_{\rm proc}^{\rm c}$ and $C_{\rm proc}^{\rm s}$ are the processing resource utilization due to communication and sensing tasks, respectively\footnote{In this paper, we focus on the GOPS analysis by taking into account only physical-layer communication and sensing processing and neglect high-layer operations.}.  

  The computational complexity is assessed by counting the numbers of real multiplications and divisions, where each complex multiplication is equivalent to four real multiplications. We also consider memory overhead in arithmetic operation calculations by multiplying each operation by two as done in \cite{desset2016massive,demir2023cell}. Hence, each complex multiplication is counted as $4\cdot 2= 8$ operations in computing the total GOPS. 
\vspace{-3mm}
\subsection{Communication GOPS Analysis}
We analyze the GOPS for digital signal processing corresponding to the communication tasks including the uplink channel estimation and downlink transmission. To compute the number of real multiplications, we mainly follow the GOPS analysis in \cite[App. B]{bjornson2017massive}.

We employ the LMMSE channel estimation approach from \cite{wang2020uplink}. 
Depending on the number of UEs and number of orthogonal pilots $L_p$, the computational complexity of channel estimation for all APs, denoted by $C_{\textrm{ch-est}}$, is equal to 
\begin{align}
  C_{\textrm{ch-est}}\! =\!\left\{\begin{matrix}
 8M L_p N_{\rm ue}N_{\rm tx}+8 M^2 N_{\rm ue}N_{\rm tx}, & {L_p\geq N_{\rm ue}}\\{8M L_p^2 N_{\rm tx}+8M^2N_{\rm ue}N_{\rm tx}}, &{L_p< N_{\rm ue}}. 
\end{matrix}\right. 
\end{align}
where the first term corresponds to multiplying the received uplink pilot signals by the conjugate pilot sequences, and the second term represents the computation of the LMMSE channel estimates.
It is worth noting that computing the covariance matrices of the estimated channels and estimation errors is also required. But since these statistics remain constant over many coherence blocks, the corresponding pre-computation overhead can be neglected.

 The number of real multiplications/divisions to compute centralized RZF precoding vectors for all UEs from~\eqref{wi} is 
\begin{align}
 &C_{\textrm{prec-comp}} =
  \underbrace{\frac{8\!\left((MN_{\rm tx})^2+MN_{\rm tx}\!\right)}{2}N_{\rm ue}}_{\textrm{inside the parentheses}} \!+ \!\underbrace{\frac{8\!\left((M\!N_{\rm tx})^3\!-\!MN_{\rm tx}\right)}{3}}_{\rm inversion}\nonumber\\
  &\hspace{3mm}+\! N_{\rm ue}\cdot\underbrace{8(MN_{\rm tx})^2}_{\textrm{for}(.)^{-1}\hat{\textbf{h}}_i}\!+\!  N_{\rm ue}\cdot\left(\underbrace{8MN_{\rm tx}}_{\textrm{computing }||\textbf{w}_i||} \!+\!\underbrace{4MN_{\rm tx}}_{\textrm{normalization}}\right)\nonumber\\
  & \hspace{3mm}= 12(MN_{\rm tx})^2N_{\rm ue}+ 16MN_{\rm tx} N_{\rm ue}+\frac{8(MN_{\rm tx})^3-8MN_{\rm tx}}{3}.
\end{align}

Reciprocity calibration and multiplication of the symbols by the precoding vectors, each costs $8L_dMN_{\rm ue}N_{\rm tx}$ real operations \cite{malkowsky2017world, demir2023cell}. Multiplying by the power coefficients also costs $4L_dMN_{\rm ue}N_{\rm tx}$. Finally, the GOPS corresponding to communication processing 
 (i.e., channel estimation, precoding and reciprocity calibration) is computed as
\begin{align}\label{eq:C_proc}
    &C_{\textrm{proc}}^{\rm c}\! =\! \frac{B}{L10^9} \Bigg(  C_{\textrm{ch-est}}+C_{\textrm{prec-comp}}+ 20L_dMN_{\rm ue}N_{\rm tx} \Bigg).
\end{align}
where we divided the total giga operations by the duration of the transmission block $L/B$.
\begin{table*}[!t]
\caption{Computational Complexity}
    \centering
    \begin{tabular}{|c|c|c|}
    \hline
\multirow{1.5}{*}{ \text{Notation}}& \multirow{1.5}{*}{ \text{Operation}}& \multirow{1.5}{*}{ \text{Computational complexity}}\\[1.5mm]
 \hline
        \multirow{1.5}{*}{$C_G$}& \multirow{1.5}{*}{ $\textbf{G}[m]$}& \multirow{1.5}{*}{ $20\, M N_{\rm tx}N_{\rm rx}$ } \\[1.5mm]
         \multirow{1.5}{*}{$C_a$}&\multirow{1.5}{*}{ $\textbf{G}^H[m] \textbf{y}[m]$}& \multirow{1.5}{*}{ $8\,M^2\,N_{\rm rx}\,N_{\rm tx}\,$}  \\[1.5mm]
         \multirow{1.5}{*}{$C_b$}&\multirow{1.5}{*}{ $\textbf{X}^H[m] \textbf{y}[m]$}&  \multirow{1.5}{*}{ $8\,M^2\,N_{\rm rx}\,N_{\rm tx}$ }\\[1.5mm]
         \multirow{1.5}{*}{$C_C$}&\multirow{1.5}{*}{ $\textbf{G}^H[m]\textbf{G}[m]$}&\multirow{1.5}{*}{  $ 4\,M N_{\rm rx} \left(N_{\rm tx}^2+N_{\rm tx}\right)$ }\\[1.5mm]
         \multirow{1.5}{*}{$C_D$}&\multirow{1.5}{*}{ $\textbf{X}^H[m]\textbf{X}[m]$}& \multirow{1.5}{*}{ $4\left((MN_{\rm tx})^2+MN_{\rm tx}\right)$}\\[1.5mm]
         \multirow{1.5}{*}{$C_E$}&\multirow{1.5}{*}{ $\textbf{G}^H[m]\textbf{X}[m]$}& \multirow{1.5}{*}{ $8\,M^2\,N_{\rm rx}\,N_{\rm tx}^2$}   \\[1.5mm]
        \multirow{1.5}{*}{$C_{\rm  invC}$}&\multirow{1.5}{*}{ $\textbf{C}^{-1}$ }& \multirow{1.5}{*}{ $8\frac{(N_{\rm tx}^3-N_{\rm tx})N_{\rm  rx}}{3}$} \\[1.5mm]
        \multirow{1.5}{*}{$C_{\rm  invD}$}&\multirow{1.5}{*}{ $\textbf{D}^{-1}$ }& \multirow{1.5}{*}{ $\frac{8}{3}\left((M^2N_{\rm tx}N_{\rm rx})^3-M^2N_{\rm tx}N_{\rm rx}\right)$} \\[1.5mm]
         \multirow{1.5}{*}{$C_{\rm inv}$}&$\begin{bmatrix}
        \textbf{C}& \textbf{E} \\
        \textbf{E}^H&\textbf{D} 
        \end{bmatrix}^{-1}$& \makecell{ $\frac{8}{3}
        \left(\left((1+M^2)N_{\rm tx}N_{\rm rx}\right)^3-(1+M^2)N_{\rm tx}N_{\rm rx}\right)$}  \\[3mm]
        \multirow{1.5}{*}{$C_{T_{\rm c-unaware}}$}&\multirow{1.5}{*}{$\textbf{a}^H \textbf{C}^{-1}\textbf{a}$} &\multirow{1.5}{*}{ $8\left((N_{\rm tx}N_{\rm rx})^2+N_{\rm tx}N_{\rm rx}\right)$}\\[3mm]
        \multirow{1.5}{*}{$C_{T_{\rm c-aware}}$}&\makecell{$\begin{bmatrix}
\textbf{a}^H \!&
\textbf{b}^H
\end{bmatrix}\! \begin{bmatrix}
(.) & (.) \\ 
(.)& (.)
\end{bmatrix}\begin{bmatrix}
\textbf{a}\\ 
\textbf{b}
\end{bmatrix}$}&\multirow{1.5}{*}{ $8\left((1+M^2)N_{\rm tx}N_{\rm rx}\right)^2+8(1+M^2)N_{\rm tx}N_{\rm rx}$}\\[3mm]
\hline

    \end{tabular}
    
    \label{tab:C_sensing}\vspace{-6mm}
\end{table*}

\vspace{-4mm}
\subsection{Sensing GOPS Analysis}
The GOPS analysis for sensing operations includes both signal transmission and processing the received signal. The GOPS associated with sensing transmission comprises the computation of the sensing precoding vector $\textbf{w}_0$, denoted by $C_{\rm prec-comp}^{\rm s}$, as well as obtaining the sensing signals by multiplying sensing symbols with  $\textbf{w}_0$ and power coefficient $\rho_0$.
The latter operation requires $12L_dMN_{\rm tx}$ real multiplications/divisions.  
The unitary matrix $\textbf{U}$ in \eqref{eq:w0} (subspace spanned by the UE channel estimation vectors) is already obtained when computing the RZF precoding vectors by matrix inversion and the corresponding $\mathbf{L}\mathbf{D}\mathbf{L}^H$ decomposition \cite[Lem.~B.2]{bjornson2017massive}. Therefore, the computational complexity of the ZF precoding vector $\textbf{w}_0$ is given by
\begin{align}
    C_{\rm prec-comp}^{\rm s} &=
  8(MN_{\rm tx})^2+ 12MN_{\rm tx},
\end{align}
 where the first term stands for matrix-vector multiplication and the second term corresponds to the cost of computing $||\textbf{w}_0||$ and normalization, which are counted as $8MN_{\rm tx}+4MN_{\rm tx}$.
 
After downlink transmission, the sensing receive APs forward their received signals to the cloud for processing. Let $C_{\rm se-comp}$ denote the computational complexity at the cloud required to process these signals and obtain the detector’s test statistic, expressed as
\begin{align}\label{eq:C_SP}
    C_{\rm se-comp} &= L_d C_{\rm se-prep} + C_{\rm se-detector},
\end{align}
where $C_{\rm se-prep}$ is a blocklength-dependent pre-processing term for computing the vectors and matrices in \eqref{eq:T_SP} and \eqref{eq:T}, and $C_{\rm se-detector}$ is a blocklength-independent term corresponding to matrix inversions and the final computation of the test statistic. 

Table~\ref{tab:C_sensing} summarizes the computational complexity of each component involved in forming the test statistics for clutter-aware and clutter-unaware detectors (Section~\ref{sec:detection}), leveraging the block-diagonal structure of $\textbf{C}$ to simplify the matrix inversion. 
Depending on the detector, the terms in \eqref{eq:C_SP} are obtained as
\begin{align}
&C_{\rm se\text{-}prep} \!=\!
\begin{cases}
C_G + C_a + C_C, & \hspace{-5mm}\text{clutter-unaware}, \\
C_G + C_a + C_b + C_C + C_D + C_E, & \hspace{-1mm}\text{clutter-aware},
\end{cases} \\
&C_{\rm se\text{-}detector}\!=\!
\begin{cases}
C_{\rm invC} + C_{T_{\rm c\text{-}unaware}}, & \text{clutter-unaware}, \\
C_{\rm invD} + C_{\rm inv} + C_{T_{\rm c\text{-}aware}}, & \text{clutter-aware}.
\end{cases}
\end{align}
Finally, the total sensing GOPS is
\begin{align}\label{eq:c_proc_s}
    C_{\rm proc}^{\rm s} =  \frac{B}{L10^9} &\Bigg(12L_dMN_{\rm tx}+C_{\textrm{prec-comp}}^{\rm s}+L_d C_{\rm se-prep} \nonumber\\
    &+ C_{\rm se-detector}\Bigg).
\end{align}
Note that since the detection threshold is assumed fixed, its computation is omitted from the complexity analysis.
\subsection{Total Energy Consumption}
The total energy required to complete communication and sensing tasks over one transmission block can be formulated as a function of blocklength and power coefficients as follows:
\begin{align}\label{eq:E_total}
    E_{\rm total}\!=\!\underbrace{\frac{L-L_p}{B}\Delta ^{\rm tr} \Vert \boldsymbol{\rho} \Vert^2+\frac{L}{B} E_1 \!+ \!\frac{L-L_p}{B}E_2 }_{\text{Blocklength-dependent}}\!+\underbrace{E_{\rm c}}_{\text{Constant}}
\end{align}
where 
\begin{align}
    E_c &= \frac{\Delta_{\rm cloud}^{\rm proc}}{\sigma_{\rm cool}C_{\rm max}\times10^9}\nonumber\\
    &\times \left(C_{\textrm{ch-est}}+C_{\textrm{prec-comp}}+ C_{\textrm{prec-comp}}^{\rm s}
    +C_{\rm se-detector}\right),\\
    E_1 &= \sum_{k=1}^{N_{\rm tx}}P_{\textrm{AP},0}^{\rm tx}+ \sum_{r=1}^{N_{\rm rx}} P_{\textrm{AP},0}^{\rm rx}+ P_{\rm fixed}+N_{\rm GPP}\frac{P_{\rm cloud,0}^{\rm proc}}{\sigma_{\rm cool}},\\
    E_2&=\frac{\Delta_{\rm cloud}^{\rm proc}B}{\sigma_{\rm cool}C_{\rm max}\times10^9}\nonumber\\
    &\quad \times\left(20MN_{\rm ue}N_{\rm tx}+12MN_{\rm tx}+C_{{\rm se-prep}}\right).
\end{align}
$E_1$ is the energy consumption coefficient corresponding to the fixed power terms at both radio and cloud, $E_2$ corresponds to the power terms depending on $L_d=L-L_p$, and  $E_c$ is the constant energy consumption independent of the blocklength and power coefficients
\vspace{-2mm}
\section{Joint blocklength and Power Optimization }\label{sec:optimization}
We aim to minimize the total energy consumption, including both radio and processing, by jointly optimizing the blocklength and the power coefficients while satisfying the URLLC and sensing requirements.
The optimization problem is formulated as follows:
\begin{subequations} 
\begin{align} 
\underset{\boldsymbol{\rho}\geq \textbf{0}, L>L_p}{\textrm{minimize}} \quad & \left(\frac{L-L_p}{B}\Delta ^{\rm tr} \Vert \boldsymbol{\rho} \Vert^2+\frac{L}{B} E_1 + \frac{L-L_p}{B}E_2\right) \,\label{obj_P2}
\\
\textrm{subject to} \quad &  \epsilon_i \leq \epsilon^{\rm (th)}_i, \quad \quad  i=1, \cdots, N_{\rm ue} \label{cons_reliability1}\\
& D_i^{\rm t} \leq D_i^{(\rm th)}
   ,\quad i=1, \cdots, N_{\rm ue}\label{con_L1}\\
    &D^{\rm p,c}\leq \frac{L}{B},\quad\label{con_processing_delay1}\\
   & R_{\rm s} \geq R_{\rm s}^{\rm th} 
   ,\quad \label{con_Rs1}\\&\overline{\mathsf{SINR}}_{\rm s}\geq \gamma_{\rm s},\quad \label{min:conb1}\\
    & P_k \leq P_{\rm tx},\quad\quad k=1,\cdots,N_{\rm tx} \label{min:cond1}
  \end{align}
\end{subequations}
where we have removed the constant energy consumption $E_c$ from \eqref{obj_P2} in the objective function, as it does not change the optimal values. The constraints \eqref{cons_reliability1}-\eqref{con_processing_delay1} are the URLLC requirements while \eqref{con_Rs1} and \eqref{min:conb1} are the sensing requirements with refreshing rate threshold $R_{s}^{\rm th}$ and minimum sensing SINR threshold $\gamma_{\rm s}$. $P_{\rm tx}$ is the maximum transmit power per AP. We can rewrite the per-AP power constraints in \eqref{min:cond1} in second-order cone (SOC) form in terms of $\boldsymbol{\rho}$ as 
$\left \Vert \textbf{F}_k\boldsymbol{\rho}\right \Vert \leq \sqrt{P_{\rm tx}},$ , for $k=1,\ldots,N_{\rm tx}$ where $\textbf{F}_k =\mathrm{diag}\left( \sqrt{\mathbb{E}\left\{\Vert\textbf{w}_{0,k}\Vert^2\right\}}, \ldots,\sqrt{\mathbb{E}\left\{\Vert \textbf{w}_{N_{\rm ue},k}\Vert^2\right\}}\right)$. 

Using the upper bound in \eqref{upperbound}, the constraint~\eqref{cons_reliability1} is replaced by $\epsilon_{i}^{\rm (ub)} \leq \epsilon_{i}^{\rm (th)}$, since $\epsilon_i \leq\epsilon_{i}^{\rm (ub)}$. To guarantee the reliability requirement, the condition $\epsilon_{i}^{\rm (ub)} \leq \epsilon_{i}^{\rm (th)}$ must be satisfied. Consequently, the transmission delay is upper-bounded as follows
\begin{equation}\label{eq:D_th}
    D_i^{\rm t}\leq D_i^{\rm (ub)}\triangleq\frac{L}{B\left(1-\epsilon_{i}^{\rm (th)}\right)}.
\end{equation}
To guarantee the transmission delay requirement, $D_i^{\rm (ub)}\leq  D_i^{\rm (th)}$ should be satisfied. This implies that the blocklength cannot exceed $D^{\rm (th)}_i B (1-\epsilon_i^{\rm (th)})$. 
Thus, the maximum blocklength for all UEs is
\begin{equation}\label{L_max}
    L_{\rm max,c} = \min \left\{ \left\lfloor D^{\rm (th)}_i B \left(1-\epsilon_i^{\rm (th)}\right)\right\rfloor|  i=1, \ldots, N_{\rm ue}\right\},
\end{equation}
On the other hand, the maximum blocklength that can satisfy the constraint~\eqref{con_Rs1}, is 
\begin{align}
    L_{\rm max,s}= \left\lfloor\frac{B}{1+Ba_1}\left(\frac{1}{R_{\rm s}^{\rm th}}- D^{\rm p}_{\rm 2-way}+L_pa_1-a_0\right)\right\rfloor,
\end{align}
where we have defined $\frac{G_{\rm cloud}}{N_{\rm GPP}C_{\rm max}}= a_0 + a_1(L-L_p)$ with $a_0= \frac{C_{\rm ch-est}+ C_{\textrm{prec-comp}}+C_{\textrm{prec-comp}}^{\rm s}+C_{\rm se-detector}}{10^{9}N_{\rm GPP}C_{\rm max}}$ and $a_1 = \frac{ 20MN_{\rm ue}N_{\rm tx}+12MN_{\rm tx}+C_{\rm se-prep} }{10^{9}N_{\rm GPP}C_{\rm max}}$.
Finally, the maximum tolerable blocklength satisfying both constraints is $L_{\rm max}= \min\{L_{\rm max,c}, L_{\rm max,s}\}$.

The processing delay constraint $D^{\rm p,c}\leq T$ implies that $\frac{C_{\rm cloud}}{N_{\rm GPP}C_{\rm max}}\leq 1$ must hold, which gives us a minimum acceptable blocklength as
\begin{align}
    L_{\rm min}= \max\left\{L_p+1, \left\lceil \frac{B(a_0 -a_1L_p)}{1-a_1B}\right\rceil\right\}.
\end{align}
Thus, the constraints~\eqref{con_L1}, \eqref{con_processing_delay1}, and \eqref{con_Rs1} are satisfied if $L\in [L_{\rm min}, L_{\rm max}]$ given that $L_{\rm min}\leq L_{\rm max}$.

The optimization problem is challenging to solve due to its non-convex nature, the high coupling of variables, and the combination of both continuous and integer variables. To handle the mixed-integer nature of the problem, we first relax the blocklength variable $L$ to a continuous variable $\tilde{L}$  to obtain a tractable approximation of the original problem. After obtaining the continuous solution $\tilde{L}^\star$, The corresponding feasible integer solution is then obtained by projecting $\tilde{L}$ onto the nearest integer, i.e., by selecting 
$L \in \{\lfloor \tilde{L} \rfloor, \lceil \tilde{L} \rceil\}$
that minimizes the objective function while satisfying the reliability and refreshing-rate constraints.
The relaxed optimization problem is formulated as

\begin{subequations} \label{opt: originalP2}
\begin{align} 
    \underset{\boldsymbol{\rho}\geq \textbf{0}, \tilde{L}>L_p}{\textrm{minimize}} \quad & \left(\frac{\tilde{L}-L_p}{B}\Delta ^{\rm tr} \Vert \boldsymbol{\rho} \Vert^2+\frac{\tilde{L}}{B} E_1 + \frac{\tilde{L}-L_p}{B}E_2\right)\,\label{obj_P3}
\\
\textrm{subject to} \quad &  \epsilon^{\rm (ub)}_i \leq \epsilon^{\rm (th)}_i, \quad  \forall i>0 \label{cons_reliability}\\
   & L_{\rm min}\leq\tilde{L}\leq L_{\rm max} \label{con_L}\\
   &\overline{\mathsf{SINR}}_{\rm s}\geq \gamma_{\rm s},\quad \label{min:conb}\\
    & \left \Vert \textbf{F}_k\boldsymbol{\rho}\right \Vert \leq \sqrt{P_{\rm tx}},\quad\quad k=1,\ldots,N_{\rm tx}. \label{min:cond}
  \end{align}
\end{subequations}

The optimization problem~\eqref{opt: originalP2} is not convex due to the nonconvex objective function and the constraints~\eqref{cons_reliability} and \eqref{min:conb}. To convexify the objective function, we first define a new optimization variable, denoted by  $\bar{L}$, where 
 $\tilde{L}-L_p \leq 1/\bar{L}$. The objective function is replaced by
\begin{align}
    \mathsf{F}\triangleq  \frac{1}{B}\left( \Delta ^{\rm tr} \frac{\Vert \boldsymbol{\rho} \Vert^2}{\bar{L}}+\tilde{L} E_1 + \left(\tilde{L}-L_p\right) E_2\right). \label{eq:F}
\end{align}
Minimizing the objective function is equivalent to minimizing the convex function (quadratic-over-linear plus affine function) $\mathsf{F}$ in \eqref{eq:F}. This is because minimizing this function, at the optimal solution, leads to $\bar{L} = 1/(\tilde{L}-L_p)$.

 From \eqref{upperbound}, the reliability constraints in \eqref{cons_reliability} are written as
\begin{align}\label{cons_SE2}
\ln\left(1+\frac{\rho_i\mathsf{b}_i^2}{\sum_{j=0}^{{\rm N_{ue}}}\rho_j\mathsf{a}_{i,j}^2+\sigma_n^2}\right)\! \geq 
 \frac{Q^{-1}\left(\epsilon_i^{(\rm th)}\right)}{\sqrt{\tilde{L}-L_p}}+\frac{b_i \ln{2}}{\tilde{L}-L_p},
\end{align}
where $\overline{\mathsf{SINR}}^{\rm (dl)}_{i}$ is substituted by \eqref{sinr_i}. To handle the non-convexity of the left-hand side in \eqref{cons_SE2}, we define a new variable $\chi_i$ and use fractional programming \cite{shen2018fractional2} to write the left-hand side as
\begin{align}\label{cons_SE3}
     \ln(1+\chi_i)-\chi_i+(1+\chi_i)  \frac{\rho_i\mathsf{b}_i^2}{\sum_{j=0}^{{\rm N_{ue}}}\rho_j\mathsf{a}_{i,j}^2+\rho_i\mathsf{b}_i^2+\sigma_n^2}.
\end{align}
Moreover, to represent the upper bound to $\left(\sum_{j=0}^{{\rm N_{ue}}}\rho_j\mathsf{a}_{i,j}^2+\rho_i\mathsf{b}_i^2+\sigma_n^2\right)/(1+\chi_i)$, we introduce the optimization variable $r_i$, similarly as in \cite{demir2023cell}, for $i=1,\ldots,N_{\rm ue}$, where
\begin{align}
 \frac{\sum_{j=0}^{{\rm N_{ue}}}\rho_j\mathsf{a}_{i,j}^2+\rho_i\mathsf{b}_i^2+\sigma_n^2}{1+\chi_i}\leq r_i,
 \end{align}
which is written as a SOC constraint in \eqref{con3b}. 
We then re-cast the constraint in \eqref{cons_SE3} as
\begin{align}\label{con_ccp}
    \ln\left(1+\chi_i\right)-\chi_i+\frac{\rho_i\mathsf{b}_i^2}{r_i}\geq\frac{Q^{-1}\left(\epsilon_i^{\rm(th)}\right)}{\sqrt{\tilde{L}-L_p}}+\frac{b_i \ln{2}}{\tilde{L}-L_p}
\end{align}
which will not destroy optimality since we want to minimize $r_i$ to increase the left-hand side of \eqref{con_ccp}. 

Let $\textbf{r}$ and $\boldsymbol{\chi}$ be the collective vectors $\textbf{r}= [r_1 \ \ldots \ r_{N_{\rm ue}}]^T$ and $\boldsymbol{\chi}= [\chi_1 \ \ldots \ \chi_{N_{\rm ue}}]^T$. Then, the optimal solution  $\{\boldsymbol{\rho}^{\star}, L^{\star}\}$ of the problem given below is also an optimal solution to \eqref{opt: originalP2}:
\begin{subequations} \normalfont\label{opt: originalb}
\begin{align} 
    &\underset{\boldsymbol{\rho}, \boldsymbol{\chi}, \textbf{r}\geq \textbf{0}, \tilde{L}>L_p, \bar{L}>0}{\textrm{minimize}}  \quad  \mathsf{F}
\\
    &\textrm{subject to:} \,\quad  \tilde{L}-L_p \leq \frac{1}{\bar{L}}, \label{con1b}
\\
    &\ln\left(1+\chi_i\right)-\chi_i+\frac{\rho_i\mathsf{b}_i^2}{r_i}\geq\frac{Q^{-1}\left(\epsilon_i^{\rm(th)}\right)}{\sqrt{\tilde{L}-L_p}}+\frac{b_i \ln{2}}{\tilde{L}-L_p} \label{con2b}
\\
  & \left\lVert
\begin{bmatrix}
\sqrt{2\rho_{0}}\,\mathsf{a}_{i,0} \\
\vdots \\
\sqrt{2\rho_{N_{\rm ue}}}\,\mathsf{a}_{i,N_{\rm ue}} \\
\sqrt{2\rho_i}\,\mathsf{b}_{i} \\
\sqrt{2}\sigma_n \\
1+\chi_i \\
r_i
\end{bmatrix}
\right\rVert_2
\leq 1+\chi_i+r_i,
\quad \forall i 
,\label{con3b} \\ 
     &  \eqref{con_L}, \eqref{min:conb}, \eqref{min:cond}.\nonumber
  \end{align}
\end{subequations}

However, the optimization problem in \eqref{opt: originalb} is still not convex due to the non-convex constraints \eqref{con1b}, \eqref{con2b} and \eqref{min:conb}. The terms that destroy convexity in \eqref{con1b} and \eqref{con2b} are the convex terms $1/\bar{L}$ and $\frac{\rho_i\mathsf{b}_i^2}{r_i}$ (in terms of $\boldsymbol{\rho}$ and $\textbf{r}$) on the right-hand side of \eqref{con1b} and the left-hand side of \eqref{con2b}, respectively. 
Regarding the sensing SINR constraint~\eqref{min:conb}, we first rewrite it as
\begin{equation}\label{conb3}
    \boldsymbol{\rho}^T\left(\gamma_{\rm s} \textbf{B}_D-M\,\textbf{A}_D\,  \right)\boldsymbol{\rho}\!\leq \!-\gamma_{\rm s} MN_{\rm rx}\sigma_n^2.
\end{equation}
This constraint is not convex since $-\boldsymbol{\rho}^TM\,\textbf{A}_D\boldsymbol{\rho}$ is a concave function. 
To this end, we apply the concave-convex procedure (CCP) approach to \eqref{con1b} and \eqref{con2b}, and the FPP-SCA method \cite{mehanna2014feasible} to \eqref{conb3}. To avoid any potential infeasibility issue regarding  \eqref{conb3} during the initial iterations of the algorithm, we add a slack variable $\chi_0\geq0$ and a slack penalty $\lambda$ to the convexified problem at the initial iterations. In subsequent iterations, we set $\chi_0$ to zero if it is less than a threshold, denoted as $\chi_0 \leq \epsilon_{\chi}$.
Finally, the convex problem at the  $c$-th iteration becomes 
\begin{subequations}\label{EE optimization-main}
\begin{align}
    \label{EE optimization}
   &\underset{\boldsymbol{\rho}, \boldsymbol{\chi}, \textbf{r}\geq \textbf{0}, \tilde{L}>L_p, \bar{L}>0, \chi_0\geq 0}{\textrm{minimize}} \quad  
   \mathsf{F}+\lambda \chi_0
\\
   &\textrm{subject to} \quad   {\tilde{L}-L_p} \leq\frac{2}{\bar{L}^{(c-1)}}- \frac{\bar{L}}{\left(\bar{L}^{(c-1)}\right)^2},\label{con1c}
\\
    &\ln\left(1+\chi_i\right)-\chi_i +2 \frac{\sqrt{\rho_i}^{(c-1)}\mathsf{b}_i^2\,\sqrt{\rho_i}}{r_i^{(c-1)}}  -r_i \left( \frac{\sqrt{\rho_i}^{(c-1)}\mathsf{b}_i}{r_i^{(c-1)}}\right)^2\nonumber\\
    &\hspace{5mm}\geq\frac{Q^{-1}\left(\epsilon_i^{\rm(th)}\right)}{\sqrt{\tilde{L}-L_p}}+\frac{b_i \ln{2}}{\tilde{L}-L_p}\label{con2c}
\\
  &\gamma_{\rm s}\boldsymbol{\rho}^T\textbf{B}_D\boldsymbol{\rho}-2M\,\Re\left(\left(\boldsymbol{\rho}^{(c-1)}\right)^T \textbf{A}_D\boldsymbol{\rho}\right)\nonumber\\
    &\hspace{5mm}\leq -\gamma_{\rm s} MN_{\rm rx}\sigma_n^2-M\left(\boldsymbol{\rho}^{(c-1)}\right)^T \textbf{A}_D\boldsymbol{\rho}^{(c-1)} + \chi_0, \label{con3c}
\\
& \eqref{con3b},  \eqref{con_L},  \eqref{min:cond}\nonumber.\nonumber
\end{align}
\end{subequations}

After obtaining the continuous solution $\tilde{L}^\star$, the corresponding feasible integer blocklength is determined via a nearest-integer projection. Specifically, the integer variable $L^\star$ is obtained as
\begin{subequations}\label{opt:integer}
\begin{align}
    L^\star = \arg\min_{L} \quad & \left| L - \tilde{L}^\star \right| \\
    \textrm{subject to} \quad 
    & \epsilon^{\rm (ub)}_i \leq \epsilon^{\rm (th)}_i, \quad \forall i>0, \\
    & R_{\rm s} \geq R_{\rm s}^{\rm th}, \\
    & L \in \{\lfloor \tilde{L}^\star \rfloor, \lceil \tilde{L}^\star \rceil\}.
\end{align}
\end{subequations}
which can be solved by a simple enumeration over these two values. This procedure guarantees feasibility with respect to the reliability and refreshing-rate constraints while introducing only a negligible deviation from the relaxed optimum. Specifically, since $|\tilde{L}^\star - L^\star| \leq 0.5$, the performance loss caused by the nearest-integer projection is marginal.

The proposed joint radio and processing energy minimization algorithm, referred to as the \textit{JRP-SeURLLC} algorithm, is summarized in Algorithm~\ref{alg:FPP-SCA}.
 We empirically observed that setting $\bar{L}^{(0)} = \frac{1}{L_{\rm max}-L_p}$, $\rho_0^{(0)} = 0$  and $\sqrt{\rho_i}^{(0)} = 10^{-3}\sqrt{P_{\rm tx}/N_{\rm ue}}$ for $i>0$ yields satisfactory results provides stable convergence and yields feasible solutions for the optimization problem, particularly compared to random initialization.  
\begin{algorithm}[H]
	\caption{\textit{JRP-SeURLLC} algorithm} \label{alg:FPP-SCA}
    \begin{algorithmic}
        \STATE\textbf{Initialization:} Initialize $\rho_0^{(0)} = 0$, $\sqrt{\rho_i}^{(0)} = 10^{-3}\sqrt{P_{\rm tx}/N_{\rm ue}}$ for $i>0$, $\textbf{r}^{(0)} \geq \textbf{0}$, and blocklength $\bar{L}^{(0)} = \frac{1}{L_{\rm max} - L_p}$. Set solution accuracy parameters $\epsilon, \epsilon_{\chi} > 0$, and $\lambda > 0$. Set iteration counter $c = 0$, maximum number of iterations $c_{\rm max}$, the initial objective value $\mathsf{F}^{(0)} = \infty$ and define the improvement metric as $\Delta \mathsf{F}^{(c)} = \mathsf{F}^{(c-1)} - \mathsf{F}^{(c)}$, and $\Delta\mathsf{F}^{(0)} = \infty$.       
        \STATE $c \leftarrow c + 1$
        \WHILE {$\Delta \mathsf{F}^{(c-1)} \geq \epsilon$ \textbf{and} $c \leq c_{\rm max}$}
                \STATE \hspace{5mm}{Solve the problem~\eqref{EE optimization-main} using the previous iterates $\boldsymbol{\rho}^{(c-1)}$, $\bar{L}^{(c-1)}$, and $\textbf{r}^{(c-1)}$ as constants, and update $\boldsymbol{\rho}^{(c)}$, $\bar{L}^{(c)}$, and $\textbf{r}^{(c)}$ accordingly.}
        \IF{$\chi_0 < \epsilon_{\chi}$}
                    \STATE\hspace{5mm}Set $\chi_0 = 0$ for the next iteration.
                \ENDIF
                \STATE \hspace{5mm}$c \leftarrow c + 1$
        \ENDWHILE
    \STATE Determine the integer blocklength $L^\star$ by solving \eqref{opt:integer}, i.e., by checking the feasibility of $\lfloor \tilde{L}^\star \rfloor$ and $\lceil \tilde{L}^\star \rceil$ and selecting the feasible candidate closest to $\tilde{L}^\star$.
        \STATE \textbf{Output:} Transmit power coefficients $\boldsymbol{\rho}^{\star}=\boldsymbol{\rho}^{(c)}$ and the optimized blocklength $L^{\star}$.
	\end{algorithmic}
\end{algorithm}

\vspace{-3mm}
\subsection{Complexity Analysis}
Let $c_{\max}$ denote the maximum number of outer FPP-SCA iterations required for the convergence of Algorithm~1. The overall computational complexity is therefore proportional to the complexity of solving one convexified subproblem multiplied by $c_{\max}$.
The problem is a convex program solved using an interior-point method. However, due to constraint~\eqref{con2c}, it is not a pure SOC programming.

The convexified problem in \eqref{EE optimization-main} contains $3N_{\rm ue}+4$ scalar variables. Constraint~\eqref{con2c} involves a logarithmic term which can be represented using exponential-cone programming by introducing an auxiliary variable $t_i$ such that $t_i \le \ln(1+\chi_i)$, which is equivalently written as $(1+\chi_i,1,t_i)\in\mathcal{K}_{\exp}$, where $\mathcal{K}_{\exp}$ denotes the three-dimensional exponential cone. Since each UE introduces only a constant number of auxiliary variables, the total number of variables scales linearly with the number of users, i.e., $n=\mathcal{O}(N_{\rm ue})$.

The convexified problem further contains $N_{\rm ue}$ SOC constraints from~\eqref{con3b}, $N_{\rm tx}$ SOC constraints from~\eqref{min:cond}, and one SOC-representable quadratic constraint from~\eqref{con3c}. Let $\nu$ denote the barrier parameter of the resulting conic problem, which is the sum of the barrier parameters of all cone blocks. Hence, $\nu$ scales with the total size of the SOC cones and the number of exponential-cone constraints, i.e., $\nu=\mathcal{O}\!\left(\sum_{j=1}^{N_{\rm ue}+N_{\rm tx}+1} d_j + N_{\rm ue}\right)$
where $d_j$ is the dimension of the $j$-th SOC constraint.

Using a primal-dual interior-point method, the complexity of solving one convexified subproblem is
$\mathcal{C}_{\rm SCA}
=\mathcal{O}\!\left(\sqrt{\nu}\,N_{\rm ue}^3\log\frac{1}{\epsilon}\right)$
where $\epsilon$ denotes the solution accuracy \cite{nesterov1994interior}. Therefore, the overall complexity of Algorithm~1 becomes $\mathcal{O}\!\left(c_{\max}\sqrt{\nu}\,N_{\rm ue}^3\log\frac{1}{\epsilon}\right)$.

The second subproblem performs an integer projection of the relaxed blocklength onto the feasible set $L \in \{\lfloor \tilde{L}^\star \rfloor, \lceil \tilde{L}^\star \rceil\}$. Since only two candidate values are examined, the feasibility of these two points is verified with respect to $N_{\rm ue}$ reliability constraints and one sensing-rate constraint, resulting in $\mathcal{O}(N_{\rm ue})$ complexity. This step is negligible compared with the main SCA-based optimization.
Substituting the scaling of $n$ and $\nu$, the overall complexity of the
proposed algorithm is given by $\mathcal{O}\!\left(
c_{\max}\sqrt{N_{\rm ue}+N_{\rm tx}}\;
N_{\rm ue}^{3.5}
\log\frac{1}{\epsilon}
\right)$.
\vspace{-2.5mm}
\section{Numerical Results}\label{sec:results}
In this section, we present numerical results to evaluate the performance of the proposed joint blocklength and power allocation algorithm. The simulation area spans $500\,\text{m} \times 500\,\text{m}$, with the sensing target located at the center. We have either $N_{\rm rx} \!= \!1$ or $N_{\rm rx}\! =\! 2$ sensing receive APs, where the first AP is located at coordinates $(200, 250)$ and the second at $(300, 250)$. The locations are selected relatively close to the target to achieve a high channel gain. Additionally, $N_{\rm tx} = 16$ ISAC transmit APs are uniformly distributed across the area. Each AP is equipped with $M = 4$ antenna elements unless otherwise stated. The network includes $N_{\rm ue} = 8$ URLLC UEs, randomly located in the area. The downlink transmit power is set to $P_{\rm tx} = 100$\,mW, while the uplink pilot transmission power for each UE is fixed at $50$\,mW.

The large-scale fading coefficients, shadowing
parameters, probability of LOS, and the Rician factors are simulated based on the 3GPP Urban Microcell model, defined in \cite[Table B.1.2.1-1, Table B.1.2.1-2, Table B.1.2.2.1-4]{3gpp2010further}. The path losses for the Rayleigh fading target-free channels are also modeled by the 3GPP Urban Microcell model with the difference that the channel gains are multiplied by an additional scaling parameter equal to $s=0.3$ to suppress the known parts of the target-free channels due to permanent obstacles \cite{behdad2023}. The sensing channel gains are computed by the bi-static radar range equation \cite{richards2010principles}. The carrier frequency, the bandwidth, and the noise variance are set to $1.9$\,GHz, $200$\,KHz, and $-114$\,dBm, respectively. The number of pilot symbols is $L_p = 10$. The regularization parameter $\delta$ in \eqref{wi} is set to the noise variance.

The spatial correlation matrices for the communication channels are generated by using the local scattering model in \cite[Sec.~2.5.3]{cell-free-book}. The RCS of the target is modeled by the Swerling-I model with $\sigma_{\rm rcs} = 0$\,dBsm. For all the UEs, the packet size, maximum transmission delay, and DEP threshold are $b_i=256$\,bits, $D_i^{\rm (th)}=1$\,ms, and $\epsilon_i^{\rm (th)}=10^{-5}$, respectively. 
The sensing SINR threshold is $\gamma_s=0$\,dB and the false alarm probability threshold is $P_{\rm fa}=0.03$. Detection probability results are obtained empirically through Monte Carlo simulation. The refreshing rate threshold is $R_{\rm s}^{\rm th}=10$ updates per second \cite[Table 6.2-1]{3gpp_ts_22_137}, unless otherwise stated. The remaining parameters are detailed in Table~\ref{tab:parameters}, where the values are consistent with those in \cite{demir2023cell}.

\begin{table}[t]
\centering
\caption{Simulation Parameters}
\label{tab:parameters}
\setlength{\tabcolsep}{4pt}
\renewcommand{\arraystretch}{1}
\begin{tabular}{|c|c|c|c|}
\hline
\multirow{1.5}{*}{Parameter} & \multirow{1.5}{*}{Value}
& \multirow{1.5}{*}{Parameter} & \multirow{1.5}{*}{Value} \\[1.5mm]
\hline
\multirow{1.5}{*}{$\Delta^{\rm tr},\sigma_{\rm cool}$} & \multirow{1.5}{*}{4, 0.9}
& \multirow{1.5}{*}{$C_{\rm max}$} & \multirow{1.5}{*}{700.94\,GOPS} \\[1.5mm]
\hline
\multirow{1.5}{*}{$P_{\rm fixed}$} & \multirow{1.5}{*}{120\,W}
& \multirow{1.5}{*}{$P_{\textrm{AP},0}^{\rm tx},P_{\textrm{AP},0}^{\rm rx}$}
& \multirow{1.5}{*}{$6.8\cdot M$\,W} \\[1.5mm]
\hline
\multirow{1.5}{*}{$P_{\rm cloud,0}^{\rm proc}$} & \multirow{1.5}{*}{81\,W}
& \multirow{1.5}{*}{$\Delta_{\rm cloud}^{\rm proc}$} & \multirow{1.5}{*}{288\,W} \\[1.5mm]
\hline
\end{tabular}
\vspace{-2mm}
\end{table}

\paragraph*{Benchmark Algorithms} First, we compare the performance of the proposed \textit{JRP-SeURLLC} algorithm, that jointly optimizes power and blocklength, against a fixed-blocklength optimization baseline. Then, 
we compare the performance of the proposed \textit{JRP-SeURLLC} algorithm, which aims to minimize total radio and processing energy consumption while satisfying both sensing and URLLC requirements, against two benchmark schemes:
(i) \textit{Tx-SeURLLC}, which minimizes only the transmission energy consumption, and
(ii) \textit{JRP-URLLC}, which targets total energy minimization but considers only URLLC requirements. 
The baseline schemes are selected to reflect representative optimization approaches commonly adopted in the related literature.

\begin{figure}[t!]
    \centering
   \includegraphics[width=0.75\linewidth]{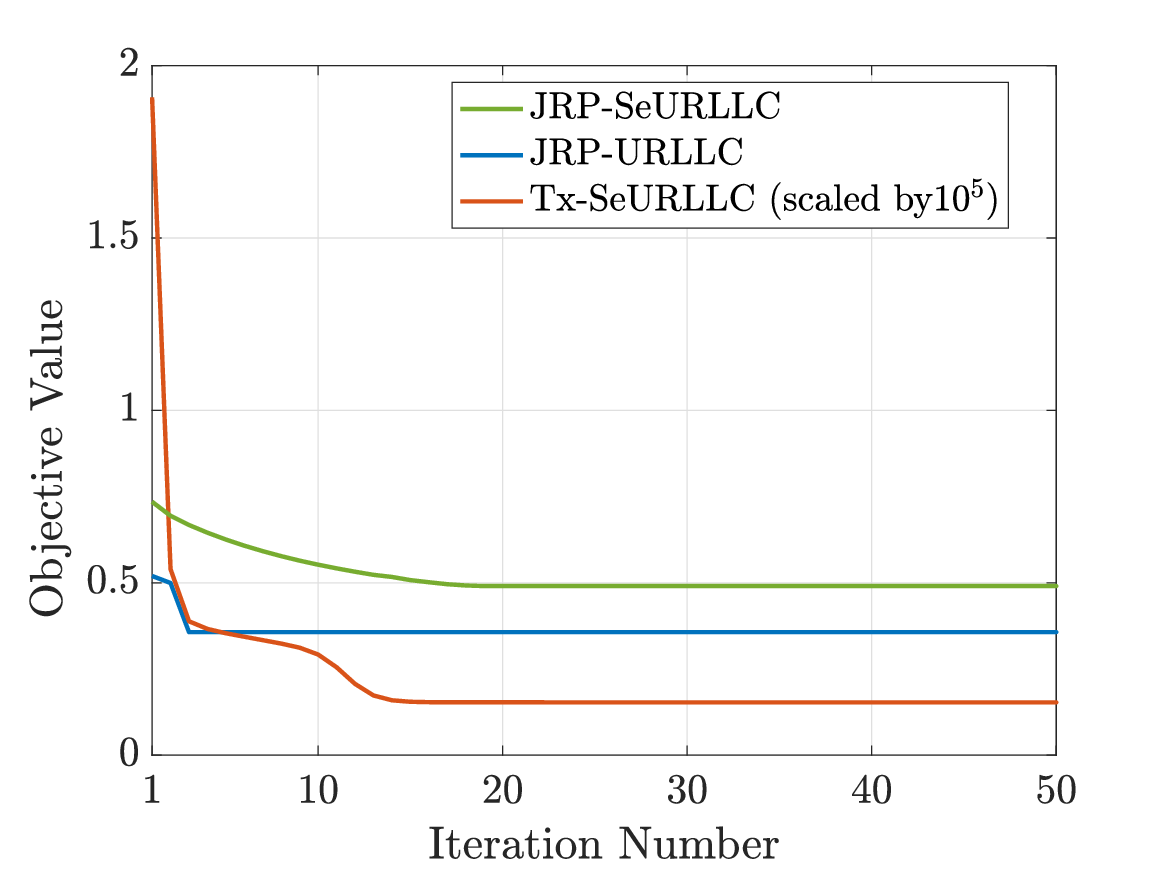} 
    \vspace{-1mm}
   \caption{Convergence condition for \textit{JRP-SeURLLC}, \textit{JRP-URLLC}, and \textit{Tx-SeURLLC} algorithms with clutter-aware detector and $N_{\rm rx}=2$. }
   \label{fig:convergence}\vspace{-3mm}
   \end{figure}

Fig.~\ref{fig:convergence} illustrates the convergence behavior of the proposed and benchmark algorithms. The values for \textit{Tx-SeURLLC} are scaled by $10^5$ since the transmission energy values are significantly lower than the total energy values. As shown, all the algorithms converge after $20$ iterations.
In the algorithms, the solution accuracy parameters are set as $\epsilon=10^{-3}$, $\epsilon_{\chi}=10^{-6}$, $\lambda = 10$, and the maximum iteration is set to $c_{\rm max}=30$. 

\begin{figure*}[t]
\centering

\subfloat[]{%
    \includegraphics[width=0.3\textwidth]{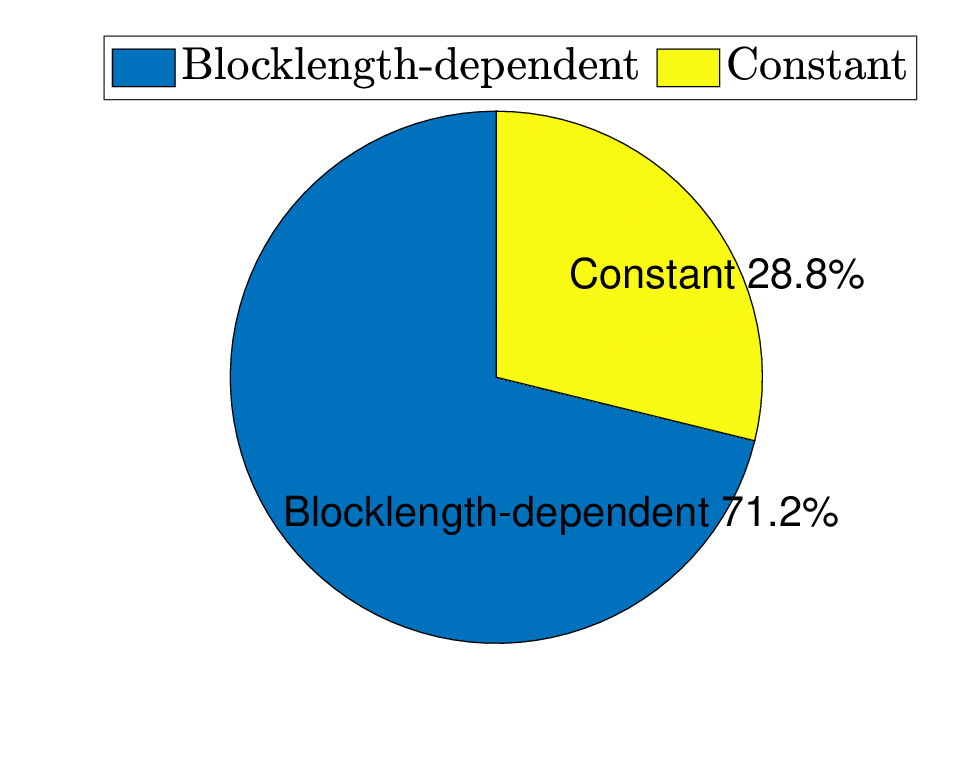}
    \label{fig: pichart}
}
\hfill
\subfloat[]{%
    \includegraphics[width=0.3\textwidth]{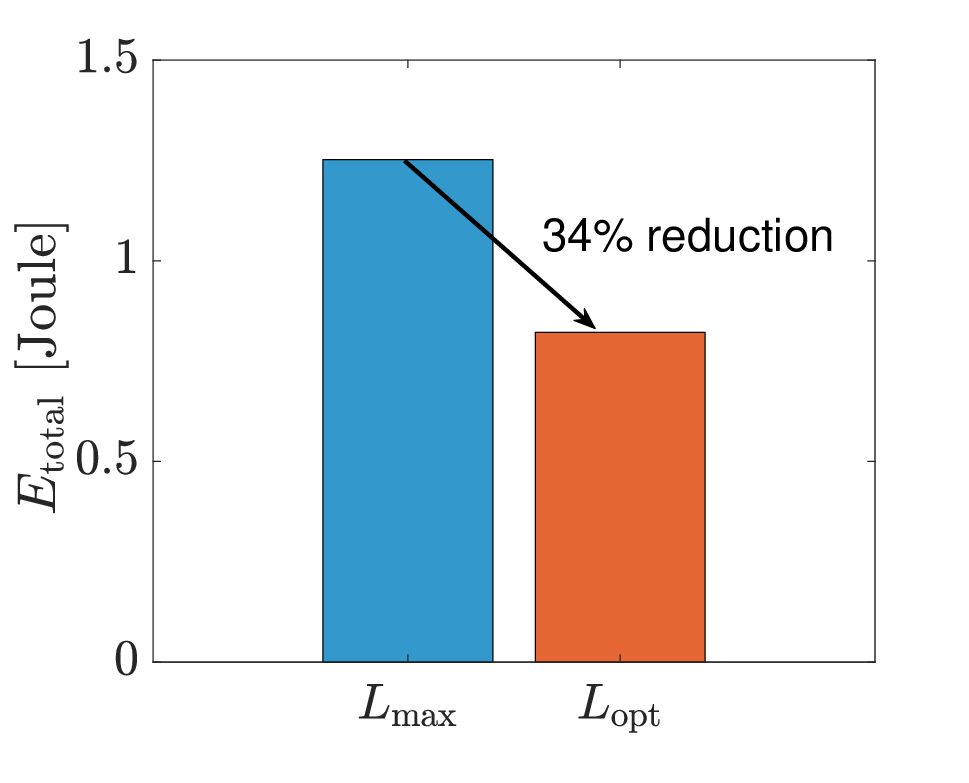}
    \label{fig: saving}
}
\hfill
\subfloat[]{%
    \includegraphics[width=0.3\textwidth]{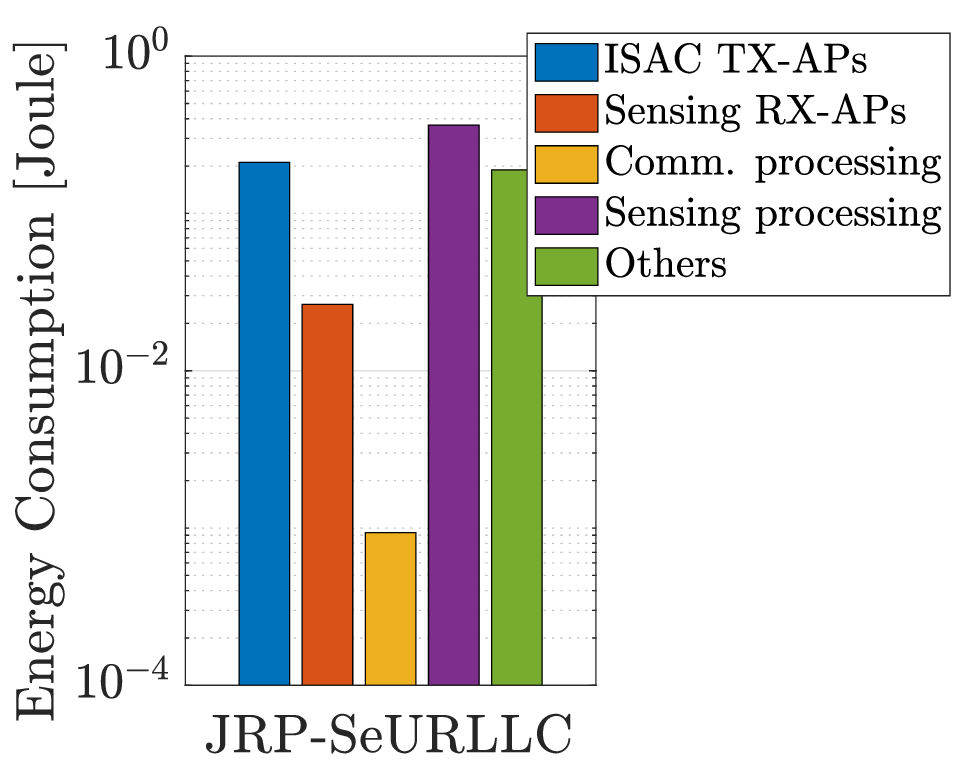}
    \label{fig: breakdown}
}
\caption{(a) Breakdown of energy consumption with clutter-aware detector, $N_{\rm rx}=2$, and $\gamma_{\rm s} = 0$\,dB, and $L=L_{\rm{max}}$, (b) total energy consumption with maximum and optimal blocklength obtained from the proposed algorithm, and (c) breakdown of the energy consumption with the proposed algorithm.}\vspace{-4mm}
\label{fig:5}
\end{figure*}

To explore the system behavior, we decompose the total energy consumption, evaluated at $L\!=\!L_{\rm max}$, into the blocklength-dependent and constant components specified in equation~\eqref{eq:E_total}. As illustrated in Fig.~\ref{fig: pichart}, approximately $71.2$\% is attributed to the blocklength-dependent terms, which is substantially higher compared to the constant energy component that accounts for only $28.8$\% of total energy. Nevertheless, Fig.~\ref{fig: saving} demonstrates that the proposed algorithm can reduce the energy consumption by up to around $34$\% through blocklength optimization compared to when maximum allowable blocklength is utilized. 

Fig.~\ref{fig: breakdown} presents a detailed breakdown of energy consumption across system components and operations, including ISAC transmit APs, sensing receive APs, communication and sensing processing, and an “Others” category representing load-independent and idle-mode power consumption in the cloud. As shown, the majority of energy consumption is attributed to sensing processing tasks, ISAC transmit APs, and the cloud's load-independent and idle-mode power usage. Notably, sensing-related processing can become a comparable or even dominant energy contributor, motivating the joint consideration of radio-side and processing-side energy consumption.

\begin{figure}[t]
        \centering
        \includegraphics[width=0.75\linewidth]{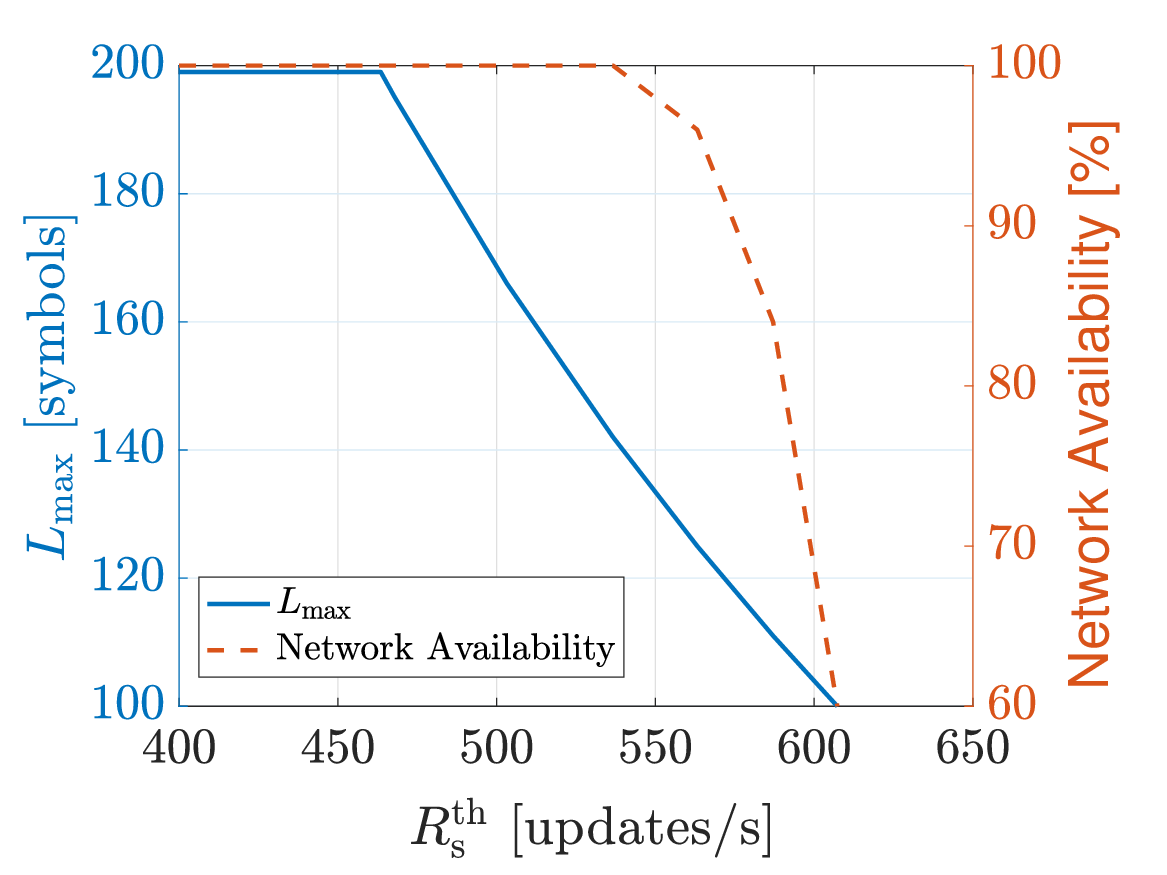} 
 \caption{Maximum blocklength threshold and network availability vs. refreshing rate threshold for $N_{\rm rx}=2$ and the \textit{JRP-SeURLLC} algorithm with 1 GPP.}   \label{fig:Lmax_Rs}\vspace{-4mm}
\end{figure}

Fig.~\ref{fig:Lmax_Rs} illustrates the maximum blocklength threshold and network availability as a function of the refreshing rate threshold. The maximum blocklength threshold $L_{\rm max}$ is limited by maximum tolerable blockength for communication, $L_{\rm max,c}$ as obtained in \eqref{L_max} and the maximum tolerable blockength for sensing, $L_{\rm max,s}$, which is a function of refreshing rate threshold. Network availability represents the percentage of cases in which the optimization problem is feasible, meaning that all requirements can be satisfied \cite{behdad2023URLLC}. The results show that, up to $R_{\rm s} \approx 460$, the maximum blocklength is limited by communication constraints. However, as the required refreshing rate exceeds $460$ updates/s, the system becomes increasingly constrained, resulting in a notable reduction in allowable blocklength. Specifically, the maximum blocklength is nearly halved when $R_{\rm s}^{\rm th} = 600$. However, such reduced blocklengths may not satisfy reliability requirements, as we observe the network availability drops to below 90\% at $R_{\rm s}^{\rm th} \approx 580$ updates/s and further declines to 60\% at 600 updates/s.

\begin{figure*}[t]
\centering

\subfloat[]{%
    \includegraphics[trim={0mm 0mm 0mm 0mm},clip,width=0.3\textwidth]{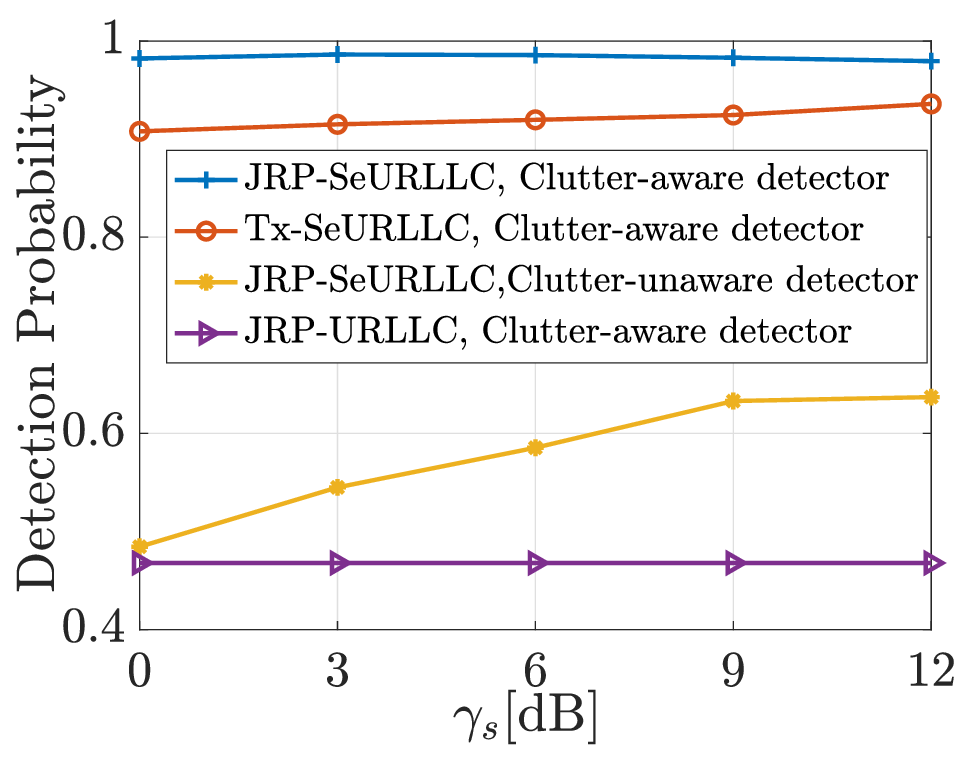}
    \label{fig:detection}
}
\hfill
\subfloat[]{%
    \includegraphics[trim={0mm 0mm 0mm 0mm},clip,width=0.3\textwidth]{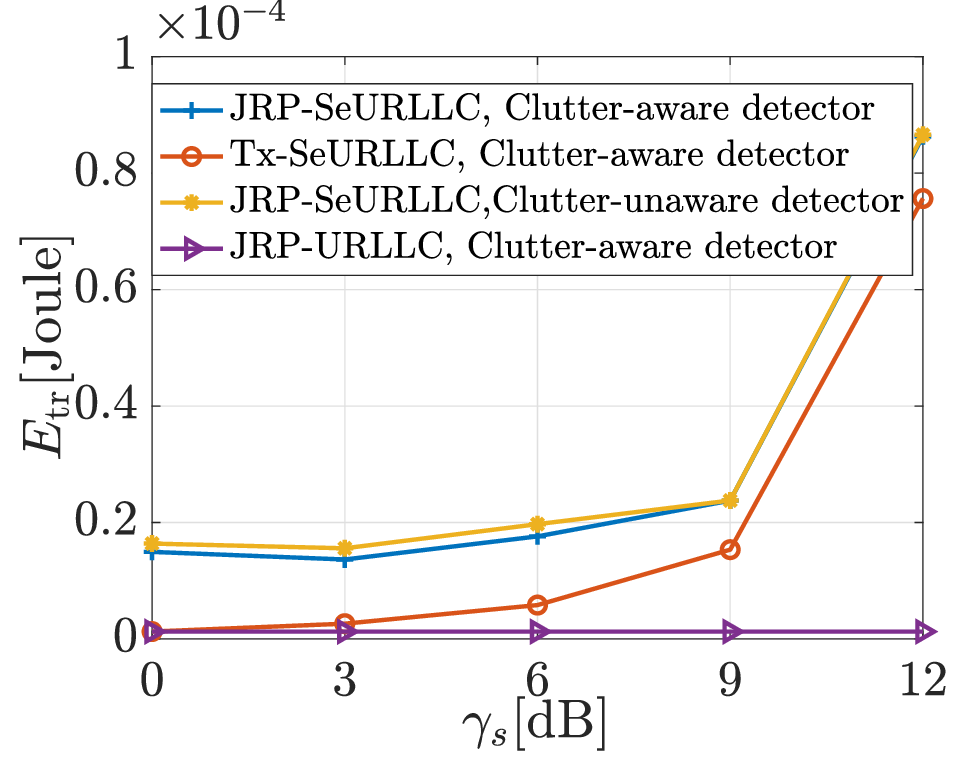}
    \label{fig:EtrvsGammaS}
}
\hfill
\subfloat[]{%
    \includegraphics[trim={0mm 0mm 0mm 0mm},clip,width=0.3\textwidth]{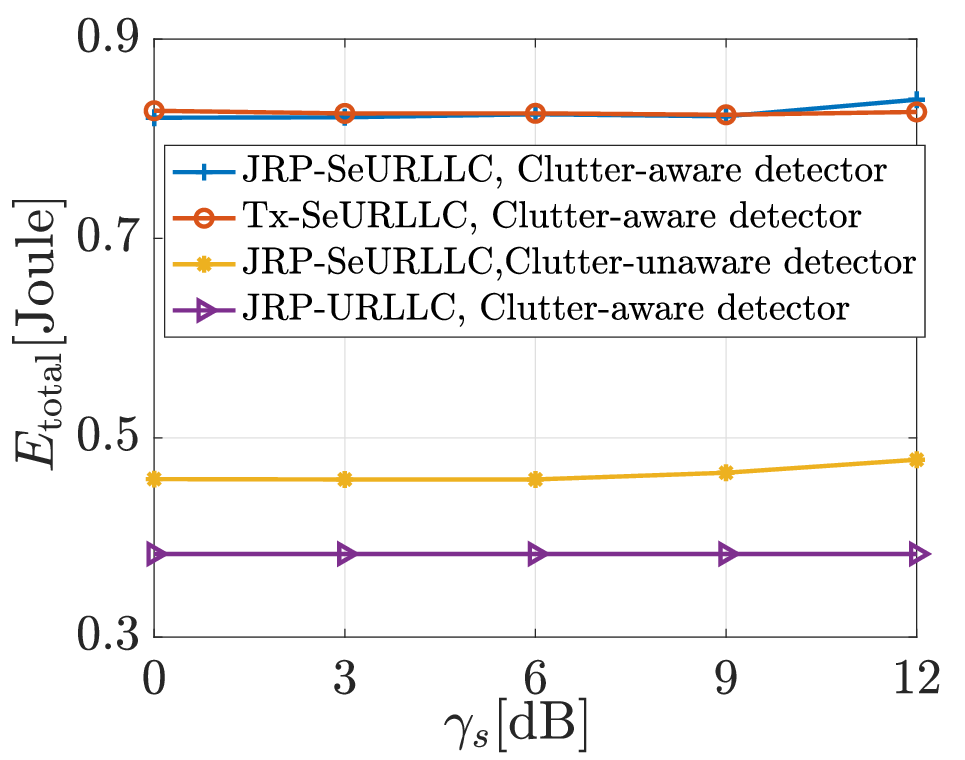}
    \label{fig:EvsGammaS}
}

\caption{(a) Detection probability, (b) transmission energy, and (c) total energy vs. sensing SINR threshold with $N_{\rm rx}=2$.}
\label{fig:gammaS}
\vspace{-4mm}
\end{figure*}

The effect of sensing SINR threshold $\gamma_{\rm s}$ on detection probability, transmission energy consumption, and total energy consumption is evaluated in Fig.~\ref{fig:gammaS}a-c, respectively. As shown in Fig.~\ref{fig:detection}, \textit{JRP-URLLC} fails to provide a detection probability above $0.5$, which highlights the need to jointly consider sensing and communication requirements in such systems. Except \textit{JRP-URLLC}, detection probability in all curves generally increases with $\gamma_s$, but performance varies based on the detector and resource allocation strategy. Clutter-aware detectors consistently outperform clutter-unaware ones due to their advanced signal processing. 
The proposed \textit{JRP-SeURLLC} algorithm with the clutter-aware detector achieves the highest detection probability up to $0.98$, while the maximum detection probability of the same algorithm with the clutter-unaware detector is $0.63$.
Moreover, with a clutter-aware detector, the \textit{JRP-SeURLLC} algorithm achieves a $4.7\%$ improvement in detection probability (from $0.935$ to $0.98$) compared to the \textit{Tx-SeURLLC} algorithm, while maintaining nearly the same total energy consumption as shown in Fig.\ref{fig:EvsGammaS}. This is because the \textit{JRP-SeURLLC} algorithm tries to minimize the total energy consumption by reducing the blocklength that allows higher transmit power levels to meet the requirements, which in turn enhances detection capability. 

Fig.~\ref{fig:EtrvsGammaS} and \ref{fig:EvsGammaS} show that in general higher $\gamma_{\rm s}$ increases transmission energy, though the rise in total energy is more moderate. Interestingly, both detectors with \textit{JRP-SeURLLC} yield similar transmission energy, while total energy consumption drops by $~43\%$ with a clutter-unaware detector—at the cost of $~35\%$ loss in detection performance. 

\begin{figure*}[t]
\centering

\subfloat[]{%
    \includegraphics[trim={0mm 0mm 0mm 0mm},clip,width=0.3\textwidth]{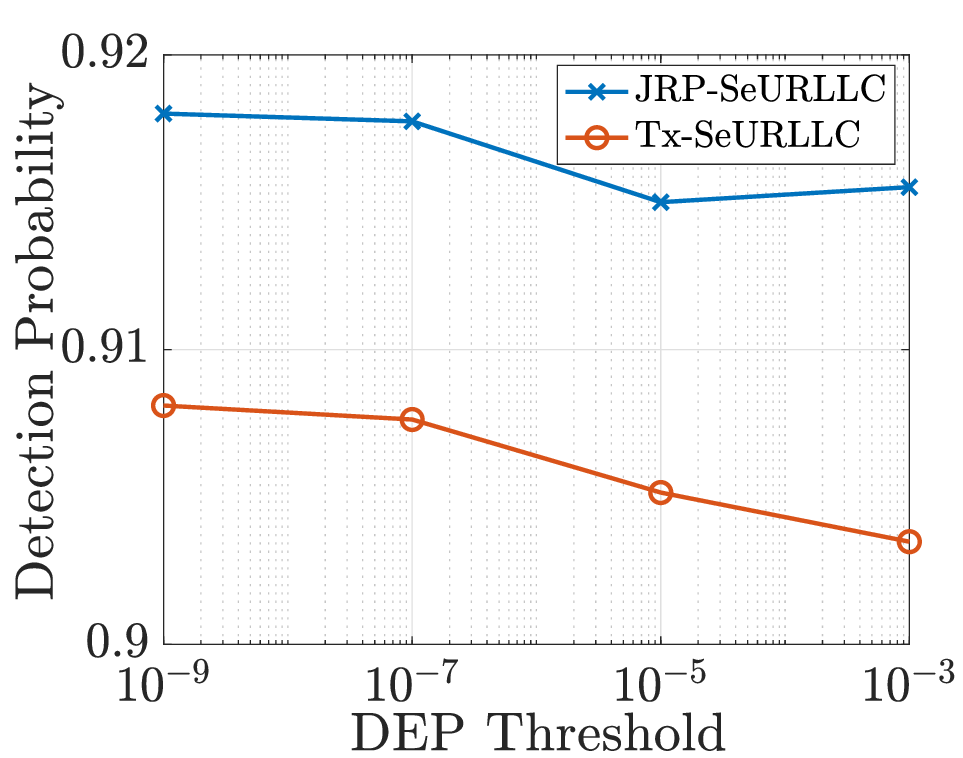}
    \label{fig:Pd_DEP}
}
\hfill
\subfloat[]{%
    \includegraphics[trim={0mm 0mm 0mm 0mm},clip,width=0.3\textwidth]{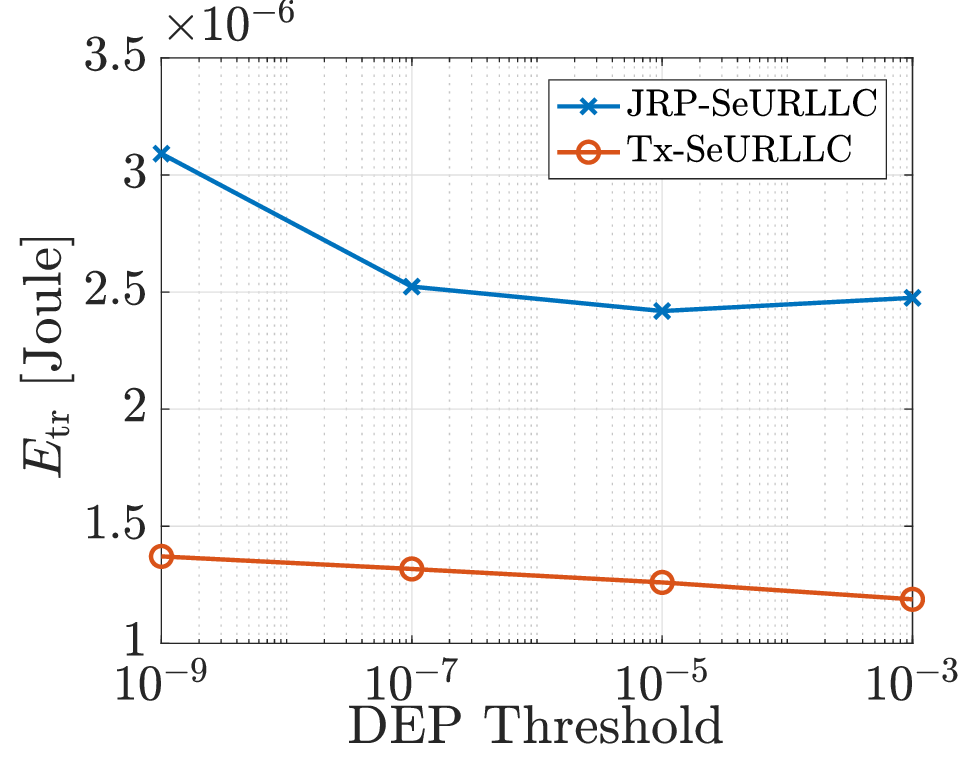}
    \label{fig:trE_DEP}
}
\hfill
\subfloat[]{%
    \includegraphics[trim={0mm 0mm 0mm 0mm},clip,width=0.3\textwidth]{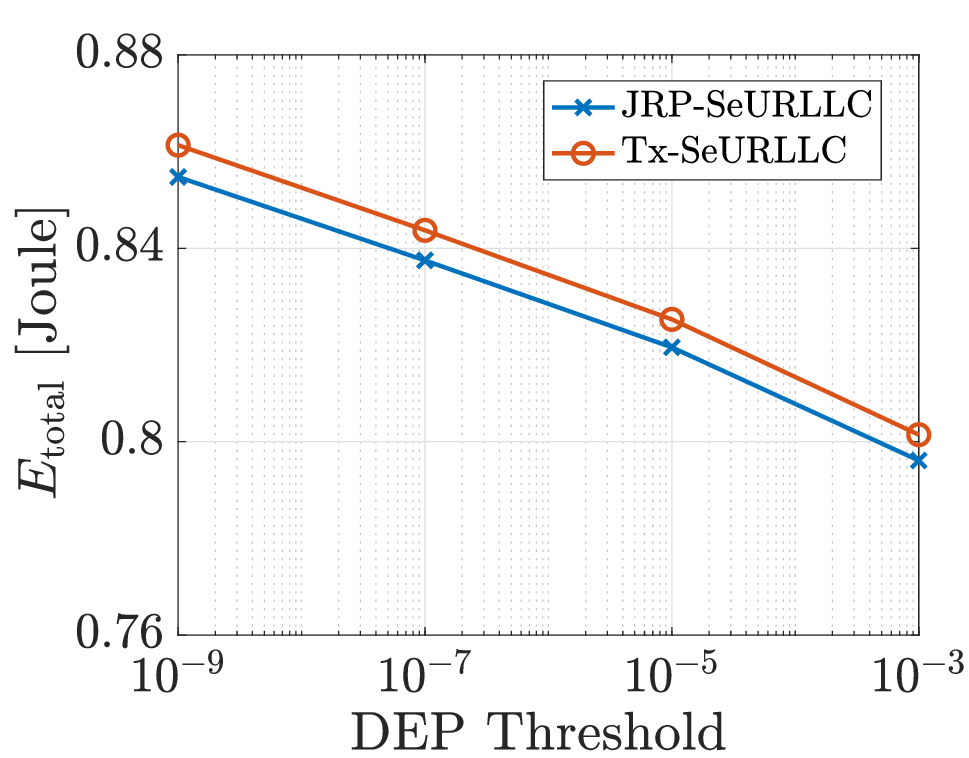}
    \label{fig:E_DEP}
}

\caption{(a) Detection probability, (b) transmission, and (c) total energy consumption vs. DEP threshold for $N_{\rm rx}=2$, $\gamma_{\rm s} = 0$\,dB.}
\label{fig:DEP}
\vspace{-4mm}
\end{figure*}

Figs.~\ref{fig:DEP}a–c illustrate the impact of the DEP threshold on sensing performance, transmission energy, and total energy consumption, respectively. 
As shown in Fig.~\ref{fig:Pd_DEP}, stricter reliability requirements slightly enhance detection probability, since the communication task requires higher power and/or longer blocklengths to meet lower DEP thresholds, which in turn enhances the sensing performance. However, this improvement comes at the cost of increased energy consumption, as evident in Fig.~\ref{fig:trE_DEP} and Fig.~\ref{fig:E_DEP}. Notably, the \textit{JRP-SeURLLC} algorithm consistently outperforms the \textit{Tx-SeURLLC} algorithm in terms of both sensing performance and energy consumption.

It is worth mentioning that a higher delay threshold allows the system to operate with a higher blocklength. However, it does not affect the results since the algorithm still chooses a smaller blocklength to minimize the energy consumption.

\begin{figure}[t!]
 \centering

\includegraphics[width=0.45\textwidth]{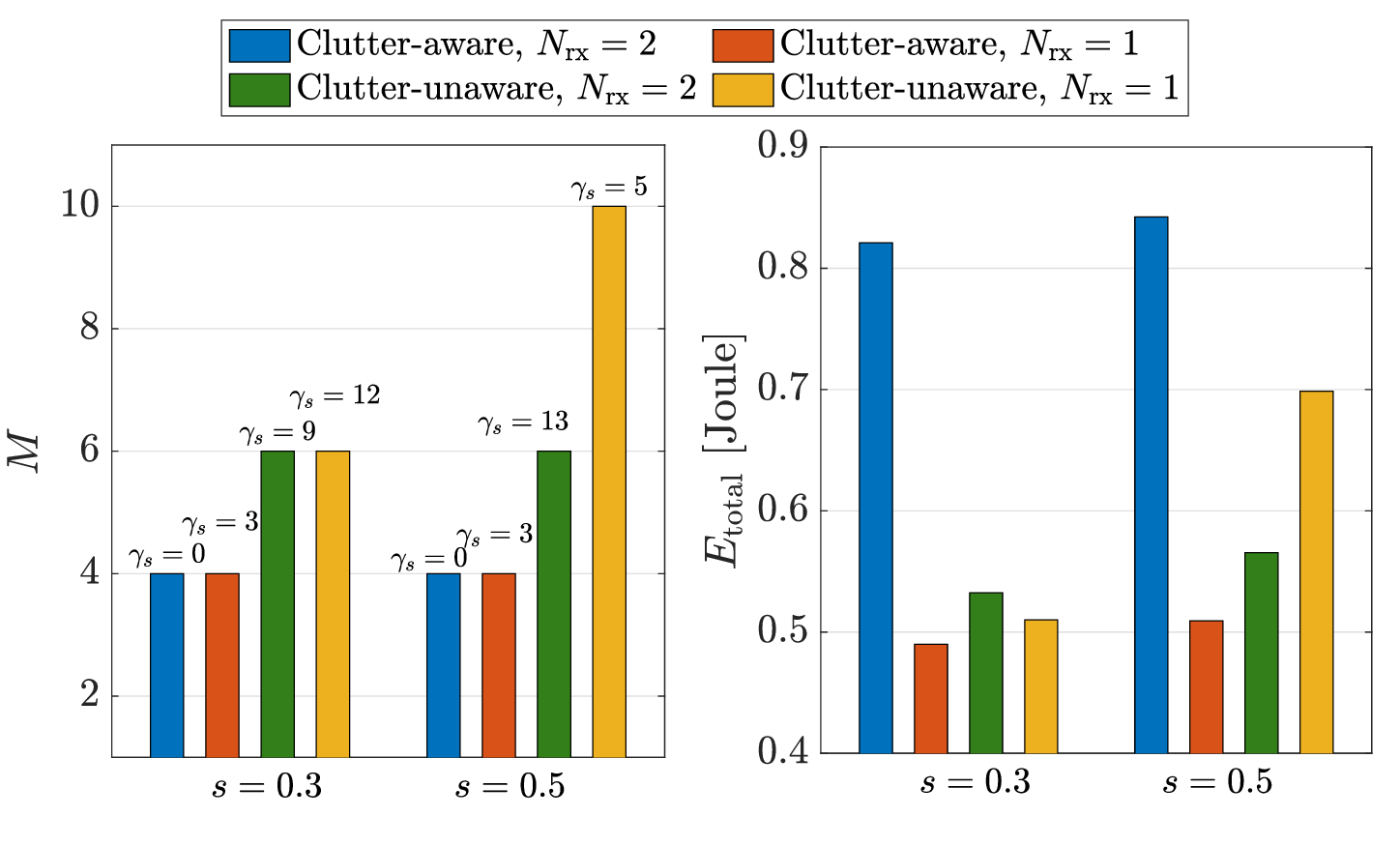}
   \put(-190,1){(a)}
   \put(-60,1){(b)}
 \caption{(a) Number of antenna elements per AP, and (b) total energy consumption for a minimum detection probability of $0.9$ with \textit{JRP-SeURLLC} algorithm and $N_{\rm rx}=1,2$.}\label{fig:6} \vspace{-6mm}
    \end{figure} 
The number of required antennas per AP and the corresponding total energy consumption to achieve a minimum detection probability of $0.9$ using the \textit{JRP-SeURLLC} algorithm under two clutter scaling parameters, $s=0.3$ and $s=0.5$, are presented in Fig.~\ref{fig:6}a and Fig.~\ref{fig:6}b, respectively. In both figures, the minimum sensing SINR thresholds that satisfy the detection requirement are indicated. The value of $\gamma_{\rm s}$ is determined from detector-specific numerical calibration curves under a fixed $P_{\rm fa}$ so as to guarantee the target detection probability. 
From Fig.~\ref{fig:6}a, achieving a detection probability of $0.9$ requires $4$ antenna elements per AP when using a clutter-aware detector, while at least $6$ antennas are needed with a clutter-unaware detector to meet the same target. With only one receive AP and $s=0.3$, both detectors require sensing SINR thresholds approximately $3$\,dB higher to maintain the desired detection probability. In addition, clutter-unaware detectors require about $9$\, dB higher sensing SINR compared to the clutter-aware ones.  For higher clutter power, i.e., $s=0.5$, the clutter-aware detector maintains the same antenna configuration and sensing SINR requirements to achieve the target detection probability. In contrast, a clear performance gap emerges for the clutter-unaware detector: it fails to meet the detection requirement under the same configuration unless a higher sensing SINR is used or additional antenna elements are deployed. 

Fig.~\ref{fig:6}b further shows that deactivating one RX-AP offers significant energy savings: up to $40\%$ with the clutter-aware detector and around $3.5\%$ with the clutter-unaware detector when $s=0.3$. Moreover, when only one RX-AP is active, the clutter-aware detector achieves slightly lower energy consumption due to its reduced antenna requirement and lower sensing SINR threshold, highlighting its efficiency advantage. 
As expected, increasing the clutter power degrades the sensing SINR and increases energy consumption to satisfy the detection constraint. Nevertheless, the main qualitative trends remain unchanged. In particular, clutter-aware detectors preserve their efficiency and robustness, whereas clutter-unaware designs require significantly higher sensing resources, either in terms of SINR or antenna elements, resulting in increased energy consumption. These results indicate that the trade-offs among energy consumption, sensing reliability, and hardware configuration persist across different clutter conditions, demonstrating the robustness of the proposed framework.

\vspace{-3mm}
\section{Conclusion and Future Directions} \label{sec:conclusion}
In this work, we proposed a joint blocklength and power control algorithm for downlink CF-mMIMO systems supporting multi-static sensing and URLLC UEs in ultra-reliable target-aware actuation use cases. We formulated a non-convex optimization problem to minimize total energy consumption, including both radio and processing energy, and analyzed two types of target detectors: clutter-aware and clutter-unaware, highlighting their respective complexity and performance trade-offs. A GOPS-based analysis was conducted for both communication and sensing tasks.  Additionally, we introduced the refreshing rate as a sensing performance metric and derived a closed-form expression that accounts for both sensing observation and processing delays.
Numerical results revealed that the sensing processing dominates overall energy consumption. The proposed algorithm achieved up to $34\%$ energy reduction compared to schemes using maximum allowable blocklength, enabling higher transmission power and shorter blocklengths to improve detection without excessive energy costs. The results also emphasize trade-offs among detector complexity, the number of antenna elements per AP, and the number of sensing receive APs. Notably, the clutter-aware detector offered significant energy savings and superior sensing performance compared to the clutter-unaware detector, albeit at increased computational complexity. Furthermore, reducing the number of active receive APs yielded up to $40\%$ energy savings with minimal impact on detection performance when the clutter-aware detector was employed. Achieving a target detection probability of $0.9$ required only four antennas per AP with a clutter-aware detector, compared to at least six antennas per AP and $9$\,dB higher sensing SINR threshold with a clutter-unaware detector.
\vspace{-4mm}
\bibliographystyle{IEEEtran}
\bibliography{IEEEabrv.bib,refs1.bib}

\end{document}